\documentclass[%
 aip,
jcp,
 amsmath,amssymb,
reprint,%
]{revtex4-2}

\usepackage{graphicx}
\usepackage{dcolumn}
\usepackage{bm}
\usepackage{amsmath}
\usepackage[utf8]{inputenc}
\usepackage[T1]{fontenc}
\usepackage{mathptmx}
\usepackage{etoolbox}
\usepackage{xcolor}
\usepackage{gensymb} 

\makeatletter
\def\@email#1#2{%
 \endgroup
 \patchcmd{\titleblock@produce}
  {\frontmatter@RRAPformat}
  {\frontmatter@RRAPformat{\produce@RRAP{*#1\href{mailto:#2}{#2}}}\frontmatter@RRAPformat}
  {}{}
}%
\makeatother

\newcommand{\resub}[1]{\textcolor{black}{#1}}

\newcommand{\um}{~\mu\mathrm{m}}

\newcommand{\figuretag}[1]{%
  \addtocounter{figure}{-1}%
  \renewcommand{\thefigure}{#1}%
}

\begin{document}

\preprint{AIP/123-QED}

\title[Controlling the volume fraction]{Controlling the volume fraction of glass-forming colloidal suspensions using thermosensitive host `mesogels'}
\author{J.S. Behra}
\affiliation{Laboratoire Charles Coulomb (L2C), Universit\'e Montpellier, CNRS, Montpellier, France}
\email{juliette.behra@umontpellier.fr}
\author{A. Thiriez}
\affiliation{Laboratoire Charles Coulomb (L2C), Universit\'e Montpellier, CNRS, Montpellier, France}
\author{D. Truzzolillo}
\affiliation{Laboratoire Charles Coulomb (L2C), Universit\'e Montpellier, CNRS, Montpellier, France}
\author{L. Ramos}
\affiliation{Laboratoire Charles Coulomb (L2C), Universit\'e Montpellier, CNRS, Montpellier, France}
\author{L. Cipelletti}%
\affiliation{Laboratoire Charles Coulomb (L2C), Universit\'e Montpellier, CNRS, Montpellier, France}
\affiliation{Institut Universitaire de France, Paris, France}
\email{luca.cipelletti@umontpellier.fr}

\date{\today}

\begin{abstract}
The key parameter controlling the glass transition of colloidal suspensions is $\varphi$, the fraction of the sample volume occupied by the particles. Unfortunately, changing $\varphi$ by varying an external parameter, \textit{e.g.} temperature $T$ as in molecular glass formers, is not possible, unless one uses thermosensitive colloidal particles, like the popular poly(\textit{N}-isopropylacrylamide) (PNiPAM) microgels. These however have several drawbacks, including high deformability, osmotic deswelling and interpenetration, which complicate their use as a model system to study the colloidal glass transition. Here, we propose a new system consisting of a colloidal suspension of non-deformable spherical silica nanoparticles, in which PNiPAM hydrogel spheres of $\sim 100-200~\mathrm{\um}$ size are suspended. These non-colloidal `mesogels' allow for controlling the sample volume effectively available to the silica nanoparticles and hence their $\varphi$, thanks to the $T$-induced change in mesogels volume. Using optical microscopy, we first show that the mesogels retain their ability to change size with $T$ when suspended in Ludox suspensions, similarly as in water. We then show that their size is independent of the sample thermal history, such that a well-defined, reversible relationship between $T$ and $\varphi$ may be established. Finally, we use space-resolved dynamic light scattering to demonstrate that, upon varying $T$, our system exhibits a broad range of dynamical behaviors across the glass transition and beyond, comparable with those exhibited by a series of distinct silica nanoparticle suspensions of various $\varphi$.
\end{abstract}

\maketitle

\section{\label{sec:intro}Introduction}

Glasses are characterized by a structure that resembles that of liquids, while microscopic dynamics are orders of magnitude slower than in fluids.~\cite{biroli_perspective_2013} Typically, the control parameter in molecular glasses is temperature, $T$: when the sample is cooled quickly enough below the glass transition temperature, 
crystallization is avoided, leading to an amorphous solid.~\cite{biroli_perspective_2013} 
Remarkably, other systems exhibit a phenomenology similar to that of molecular glass formers, \textit{e.g.} granular systems,~\cite{behringer_physics_2019} dense colloidal suspensions,~\cite{hunter_physics_2012} and active~\cite{janssen_active_2019} or biological matter.~\cite{angelini_glass-like_2011} On the one hand, these analogies motivate the quest for a general scenario for the glass transition. On the other hand, they pave the way for using systems such as colloids as model glass formers, because structural and dynamical quantities of interest are more readily accessible in colloidal suspensions than in molecular glasses.~\cite{poon_colloids_2004,hunter_physics_2012}

Colloids are sub-micron particles dispersed in a solvent. The simplest colloidal glass former comprises colloidal hard spheres, whose relevant parameter is $\varphi$, the fraction of the sample volume occupied by the particles,~\cite{pusey_phase-behavior_1986,pusey_observation_1987} rather than $T$, as in molecular glasses. In other colloidal systems, the interparticle potential is more complex than the no-overlap hard sphere potential. Both attractive and repulsive interactions are routinely encountered in colloidal systems, and in general the colloidal glass transition depends on both $\varphi$ and particle interactions. This results in a very rich behavior that often has no counterpart in molecular materials, see e.g. the non-monotonic (reentrant) glass transition of colloids with short-range attractive interactions.~\cite{pham_multiple_2002}

Investigating the colloidal glass transition usually involves preparing a series of distinct samples whose composition is varied in order to explore a range of $\varphi$ and/or of interparticle interactions. This is different from molecular systems, where a single sample may be used across the glass transition by simply varying $T$. In colloids, the lack of an easily tunable external parameter poses several challenges. Controlling and measuring the volume fraction with the required accuracy is difficult, even for hard spheres.~\cite{Poon_Weeks_Royall_2012} Manipulating concentrated colloids, \textit{e.g.} to transfer them to a measuring cell, can be quite tricky, especially for suspensions of small particles, which are in general very stiff. For example, the shear modulus of glassy hard spheres scales as $G \sim k_BT/a^{3}$, with $k_B$ Boltzmann's constant and $a$ the colloid radius.~\cite{MasonLinearViscoelasticityColloidal1995} Accordingly, a (marginally) glassy suspension of hard spheres with $\varphi \approx 0.6$ and $a=200$ nm has $G \sim 0.5$ Pa and flows easily when poured from a container, while a similar suspension with $a=20$ nm has $G\sim 500$ Pa. The latter is pasty and can only be transferred using a spatula, which inevitably introduces further uncertainties on $\varphi$. Finally, it is difficult to impose the equivalent of a well-controlled `thermal history' to a colloidal glass former, thus preventing the in-depth study of the effect of sample history on aging.~\cite{biroli_perspective_2013,hunter_physics_2012} While mechanical agitation is a popular way of initializing glassy colloidal suspensions,~\cite{viasnoff_rejuvenation_2002} a recent work suggests differences in aging when shear or a $\varphi$ quench is applied.~\cite{peng_comparison_2014} In general, a protocol alternative to mechanical agitation is desirable when studying the interplay between the microscopic structure and dynamics of colloidal suspensions and their rheological properties, an increasingly active research field.~\cite{denisov_resolving_2013,leheny_rheo-xpcs_2015,pommella_coupling_2019,edera_deformation_2021}

In response to these difficulties, the possibility of studying colloidal phase transitions by tuning the volume fraction or the interparticle interactions with $T$ has been explored for several years. One approach leverages on varying interactions, \textit{e.g.} using selectively wettable particles suspended in mixture of fluids close to its critical point,~\cite{beysens_adsorption_1985,lu_temperaturedependent_2010} or thanks to depletion forces~\cite{AsakuraInteractionparticlessuspended1958} whose strength varies with $T$.~\cite{SavageImagingSublimationDynamics2006,buzzaccaro_critical_2010} Another approach is based on micelles of self-assembled block-copolymers as colloidal objects.~\cite{alexandridis_solvent-regulated_1999} Thanks to the $T$ dependence of the affinity with the solvent of each block, it is possible to design systems where the degree of micellization, hence the colloidal volume fraction, depends on $T$.~\cite{louhichi_nucleation_2013} However, increasing the micelle number density often leads to the formation of (poly)crystalline phases, rather than glasses.~\cite{alexandridis_solvent-regulated_1999} Additionally, data interpretation is complicated by the difficulty to precisely quantify the degree of micellization as a function of $T$.

Thermosensitive microgels are sub-micrometric, deformable particles that provide another popular way to control the colloidal volume fraction. By varying $T$, the affinity of the polymer chains for the solvent is changed, resulting in microgel swelling or deswelling and hence in a change of $\varphi$ for samples at fixed microgel number density. Poly(\textit{N}-isopropylacrylamide) (PNiPAM)-based materials have been extensively studied,~\cite{mattsson_soft_2009,hunter_physics_2012,yunker_physics_2014, di_dynamics_2014, peng_comparison_2014, peng_physical_2016,li_long-term_2017,philippe_glass_2018} unveiling intriguing aspects of the glass transition of soft colloids, distinct from those of hard spheres. Features such as the `strong' (\textit{i.e.} Arrhenius-like) increase of the relaxation time on approaching the glass transition,~\cite{mattsson_soft_2009,di_dynamics_2014} supra-linear aging,~\cite{di_dynamics_2014} or the existence of high-density states where the dynamics depend surprisingly weakly on $\varphi$~\cite{li_long-term_2017,philippe_glass_2018} distinguish soft colloids from hard spheres and have become an active research field \textit{per se}.

Thermosensitive microgels, however, come with several complications. At high $\varphi$, they are subject to interpenetration with one another, shape modification, and osmotic deswelling, resulting in changes of the interparticle interactions.~\cite{li_long-term_2017, di_dynamics_2014, islam_deswelling_2019, romeo_origin_2012} It is difficult to disentangle the contribution of these phenomena from that of the variation of $\varphi$, hindering the understanding of the microscopic dynamics and rheological properties of microgel suspensions. Moreover, these phenomena make it difficult to compare experimental results to numerical simulations, where interactions are usually modeled by simplified central potentials and are assumed to be independent of particle density. Charge-stabilized hard particles such as silica colloids are an appealing alternative as model soft colloids,~\cite{philippe_glass_2018,chen_microscopic_2020} since they have a well-defined spherical shape and are not subject to osmotic swelling, interpenetration nor compression. In this case, softness arises from the shape of the screened Coulomb repulsive potential. Unfortunately, however, neither the volume fraction nor the interparticle potential of these systems can be significantly varied by tuning $T$.

To circumvent these difficulties, we develop a new colloidal system, comprising a dense suspension of charged-stabilized silica nanoparticles  and non-colloidal, thermosensitive PNiPAM spheres. The PNiPAM spheres have a typical diameter $\sim 100-200~\um$, intermediate between the sub-micron scale of usual microgels and the macroscopic scale; we thus term them `mesogels'. \resub{Since the mesogels are more than three orders of magnitude larger than the silica nanoparticles and well beyond the colloidal length scale, they do not alter the interaction potential between nanoparticles, as it would be the case if they had comparable size, \textit{e.g.} due to depletion interactions}. Furthermore, we do not expect the nanoparticles to penetrate in the PNiPAM mesogels, as their diameter ($\sim 30-40$ nm) is too large compared to the average mesh size expected for the mesogels ($\lesssim 10$ nm, see Refs. and Sec. I.A of the Supplementary Material (SM))~\cite{shibayama_small_1992, galicia_static_2009}). The role of the mesogels is to control the volume of the sample effectively available to the nanoparticles, thanks to the swelling or deswelling of PNiPAM gels upon temperature changes. Thus, the effective volume fraction of the silica nanoparticles can be simply tuned by varying $T$, paving the way for an easier sample manipulation (by transiently reducing $\varphi$), the straightforward study of $\varphi$-dependent properties with a single sample, and the investigation of the effect of an arbitrary $\varphi$ history imposed to the system.

The rest of the paper is organized as follows: in Sec.~\ref{sec:m&m} we describe the synthesis of the mesogels, as well as microscopy and Dynamic Light Scattering (DLS) setups. In Sec.~\ref{sec:results} we first present and discuss the $T$-dependent size of the mesogels, both in water and in concentrated Ludox suspensions, for various $T$ histories. We then use DLS to show that, upon the addition of a few \% vol. of mesogels, a suspension of nanoparticles
can span the whole range of dynamic behaviors from marginally supercooled to fully glassy upon changing $T$
. Finally, in Sec.~\ref{sec:conclusion} we recapitulate our main findings and briefly discuss future research paths opened by this work.

\section{\label{sec:m&m}Materials and methods}

\subsection{\label{mat}Materials}
\textit{N}-isopropylacrylamide (NiPAM) monomers ($\geq99\%$; ref: 731129-25G),
\textit{N,N'}-methylenebisacrylamide (BIS) crosslinker ($\geq99.5\%$; ref:
M7279-25G), 2-hydroxy-2-methylpropiophenone photoniator ($97\%$; ref: 405655-50ML) and LUDOX\textsuperscript{\textregistered} TM-50 colloidal silica ($50~wt\%$ suspension in H\textsubscript{2}O; ref: 420778-1L) were acquired from Sigma-Aldrich. Silicone surfactant DOWSIL\textsuperscript{TM} RSN-0749 resin ($\sim50\%$ cyclopentasiloxane, $\sim50\%$ trimethylsiloxysilicate; ref: 4119565) was provided by Dow. Silicone oil 47 V 100 (100 cSt; ref: 84542.290) and isopropanol ($\geq99.7\%$; ref: 20842.298) were bought from VWR. Diethyl ether ($\geq99.5\%$; ref: D/2450/17) was acquired from Fischer Scientific. ‘Ultrapure’ water type I was obtained from a Milli-Q\textsuperscript{\textregistered} Synergy\textsuperscript{\textregistered} - R ultrapure water station (Merck Millipore), and is next called deionized (DI) water.

\subsection{\label{met}Methods}
\subsubsection{\label{samp_prep}Sample preparation}
PNiPAM mesogel synthesis was carried out at $T_{\mathrm{room}}=20\mathrm{\degree C}$ with a similar protocol as that described by Kanai \textit{et al.}~\cite{kanai_preparation_2011} A stable water-in-silicone oil emulsion was prepared with a home-made microfluidic device (see Fig.~S1 in SM), and the so-obtained aqueous drops were subsequently polymerized under UV light. Details are provided in Sec.~I in SM. The mesogels were thoroughly washed with isopropanol prior to being transferred into DI water, and the remaining traces of silicone oil were removed with diethyl ether. The mesogels, suspended in DI water, were stored in the fridge prior to use. They looked relatively transparent to the naked eye, indicating that their structure is homogeneous upon the scale of the visible light wavelengths, consistent with previous investigations of the influence of the synthesis temperature on the structure and appearance of PNiPAM hydrogels, which showed that syntheses performed at $T$ below $25\degree$C yield more transparent PNiPAM materials as compared to syntheses performed at higher $T$~\cite{hirokawa_sponge-like_2008, hirokawa_direct_1999, kayaman_structure_1998} (see Sec.~I.A in SM).

To prepare the concentrated Ludox suspensions, $\sim20~\mathrm{g}$ of the commercial suspension (Ludox TM-50, average particle diameter 35 nm and polydispersity index 0.25 as determined by DLS on a diluted suspension) were centrifuged at $10500~\mathrm{rpm}$ during $\sim4~\mathrm{h}$ in a 3-15 centrifuge (Sigma) equipped with a 12158-H rotor (Sigma). The supernatant was then removed, and the suspensions were homogenized with a spatula. They were vortexed and further centrifuged at $\sim1500~\mathrm{rpm}$ for about $15~\mathrm{min}$ in a 2-4 centrifuge (Sigma) to release trapped air bubbles. The Ludox volume fractions $\varphi$ of all the Ludox suspensions used in the present study (including that of the commercial Ludox suspension) were determined by drying a small aliquote of the suspension, as described in the supplemental material of Ref.~\cite{philippe_glass_2018}. The concentrated Ludox suspensions were stored in the fridge prior to use.

Two types of mesogels and Ludox mixtures were prepared: suspensions of mesogels (i) in the commercial Ludox solution ($\varphi=0.350$) and (ii) in concentrated Ludox suspensions ($\varphi=0.396$ and $\varphi=0.412$ for the suspensions characterized with optical microscopy and DLS, respectively). The vial containing the mesogels was gently shaken to redisperse the mesogels which had sedimented over time. Mesogels were quickly sampled with a plastic pipette and immediately transferred into a $2.0~\mathrm{mL}$ Eppendorf tube. They were left to sediment in the tube and as much as possible DI water was removed. The Ludox suspension was then added. For concentrated Ludox suspensions, centrifugation was carried out to bring the Ludox suspension to the bottom of the Eppendorf tube. Finally, the mixture was gently mixed with a spatula to disperse the mesogels throughout the sample. Samples obtained from the concentrated Ludox suspension were centrifuged at low speeds to remove trapped air bubbles.

\subsubsection{\label{microscopy}Optical microscopy}

Using a plastic pipette for samples prepared in DI water and in commercial Ludox solution, and a spatula for those prepared in concentrated Ludox suspensions, the samples were transferred in a $130~\mathrm{\mu L}$ volume cell formed by a microscope slide and a coverslip spaced by two superimposed $65~\mathrm{\mu L}$ gene frames ($16~\mathrm{mm}~\times~15~\mathrm{mm}~\times~0.25~\mathrm{mm}$; ref: AB0577; Thermo Scientific), to ensure no mesogel is squeezed. \resub{To ease the microscopy observations, the mesogel volume fraction was kept low, about 0.045\% at room temperature}. To measure the temperature directly in the observation cell, a thermocouple connected to a temperature controller (di 32 PID; Jumo) was sandwiched between the two gene frames with the sensor being right in the middle of the cell. The microscope slide was then placed under the microscope (Laborlux 12 Pol S; Leitz) in a microscope hot stage (HS400; Instec) linked to a temperature controller (STC200; Instec) and a liquid nitrogen cooling unit (LN2-P; Instec). Micrographs were taken with a digital camera (D5200; Nikon) and processed with ImageJ (version 1.53f51; National Institutes of Health; USA). In particular, we used it to determine $d$, the Feret's diameter of the mesogels,~\cite{ImageJ_Ferret} which for our spherical mesogels effectively corresponds to the usual diameter.

Prior to any observation, samples were allowed to equilibrate in the microscope hot stage for at least $20~\mathrm{min}$. To characterize the mesogels at equilibrium at different $T$, the temperature was increased in small steps (no more than $1\mathrm{\degree C}$ at a time) using $T$ ramps of $+0.2\mathrm{\degree C/min}$. After each ramp, the mesogels were allowed to thermally equilibrate for $\sim 10~\mathrm{min}$ prior to taking pictures of the mesogels across the sample and subsequently measuring their size. 
To characterize the mesogel behavior upon successive changes in temperature, $T$ cycles were performed, with $\dot T_{\mathrm{up}}$ ranging from $+0.02\mathrm{\degree C/min}$ to $+3\mathrm{\degree C/min}$, and $\dot T_{\mathrm{down}} = -0.5\mathrm{\degree C/min}$. After each ramp up and down, the mesogels were allowed to thermally equilibrate for $\sim 10~\mathrm{min}$ and $\sim 5~\mathrm{min}$, respectively, prior to taking pictures of the mesogels across the sample and subsequently measuring their size. 

\subsubsection{\label{DLS}Dynamic light scattering}
The sample was prepared as described in Sec.~\ref{samp_prep}. A mass of $1.450~\mathrm{g}$ concentrated Ludox suspension ($\varphi = 0.412$) was added to $1.094~\mathrm{g}$ mesogels (as determined after removal of the supernatant; note that this amount also accounts for the water in-between the mesogels which we assumed to be at the random packing volume fraction). Using a thin spatula, $\sim 200~\mathrm{mg}$ of the sample were deposited on the inner walls of a NMR tube of diameter 4 mm, cut to a height of about 8 cm. The NMR tube was centrifuged to bring the sample down to its bottom and to remove trapped air bubbles. It was then placed in a hemolysis tube filled with DI water (diameter 1 cm), which in turn was placed in the temperature-controlled copper sample holder of the home-built set-up performing space-resolved DLS measurements described by El Masri \textit{et al.}~\cite{masri_dynamic_2009}. All the measurements were performed at a scattering angle of $90\mathrm{\degree}$, corresponding to a scattering wave vector with modulus $q = 22.3 \um^{-1}$. The power of the laser ($532~\mathrm{nm}$ wavelength \textit{in vacuum}) was set to $150~\mathrm{mW}$ for the highest investigated $T$ and was then decreased to $37.5~\mathrm{mW}$  for all the other measurements (see discussion). The measurements were performed from high to low $T$, \textit{i.e.} from the sample with the fastest dynamics to the slowest one.

\section{\label{sec:results}Results and discussion}

\begin{figure} 
\includegraphics[width=\columnwidth]{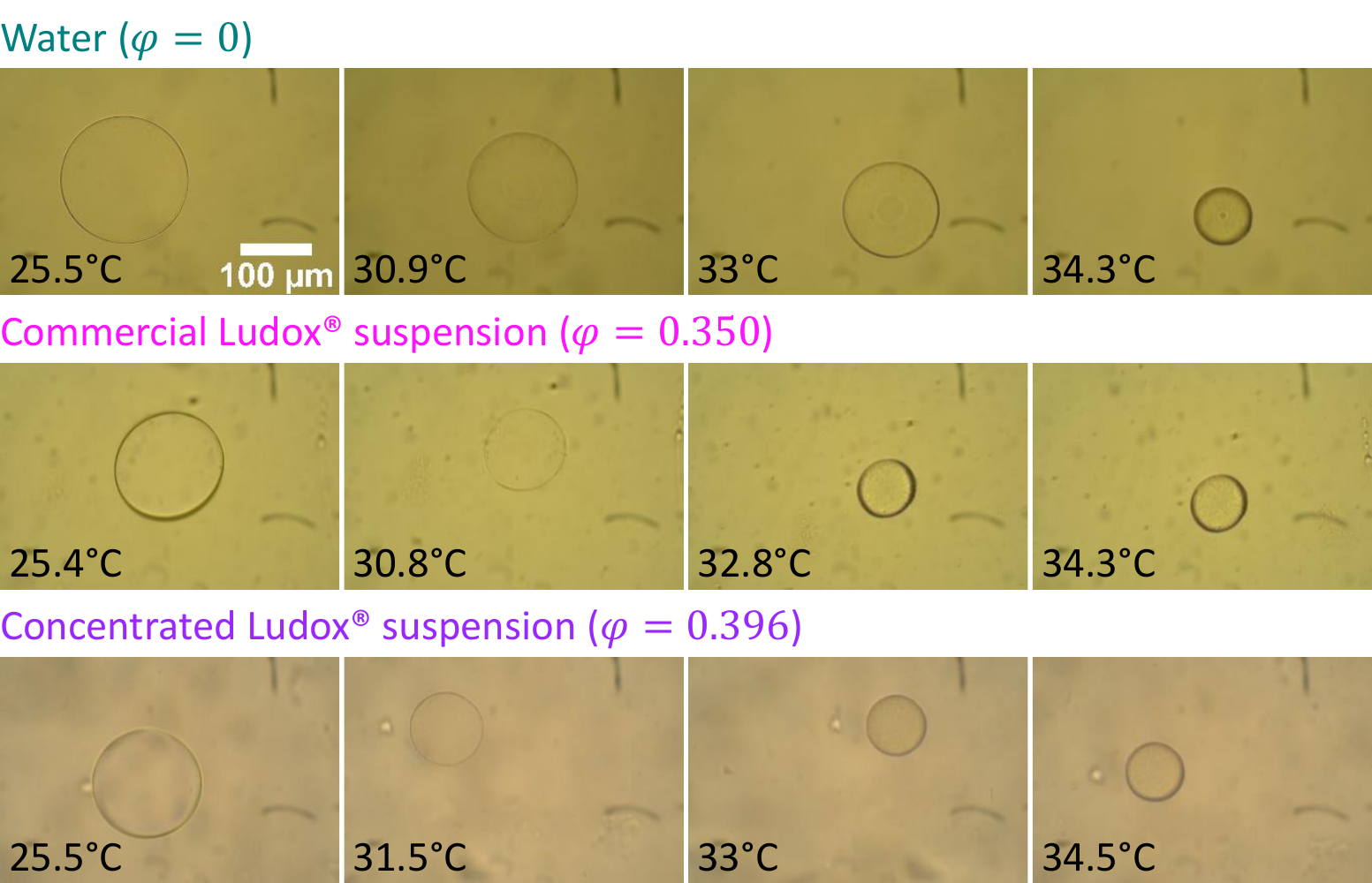}
\caption{\label{fig:ballsmicroscopy} Micrographs of equilibrated PNIPAM mesogels at different temperatures and immersion media (\textit{i.e.} pure water or Ludox suspensions with $\varphi = 0.350$ and $\varphi = 0.396$, for the top, mid and bottom row, respectively). Images were collected at magnification $\times 20$. The scale bar applies to all of them.}
\end{figure}

\begin{figure} 
\includegraphics[width=\columnwidth]{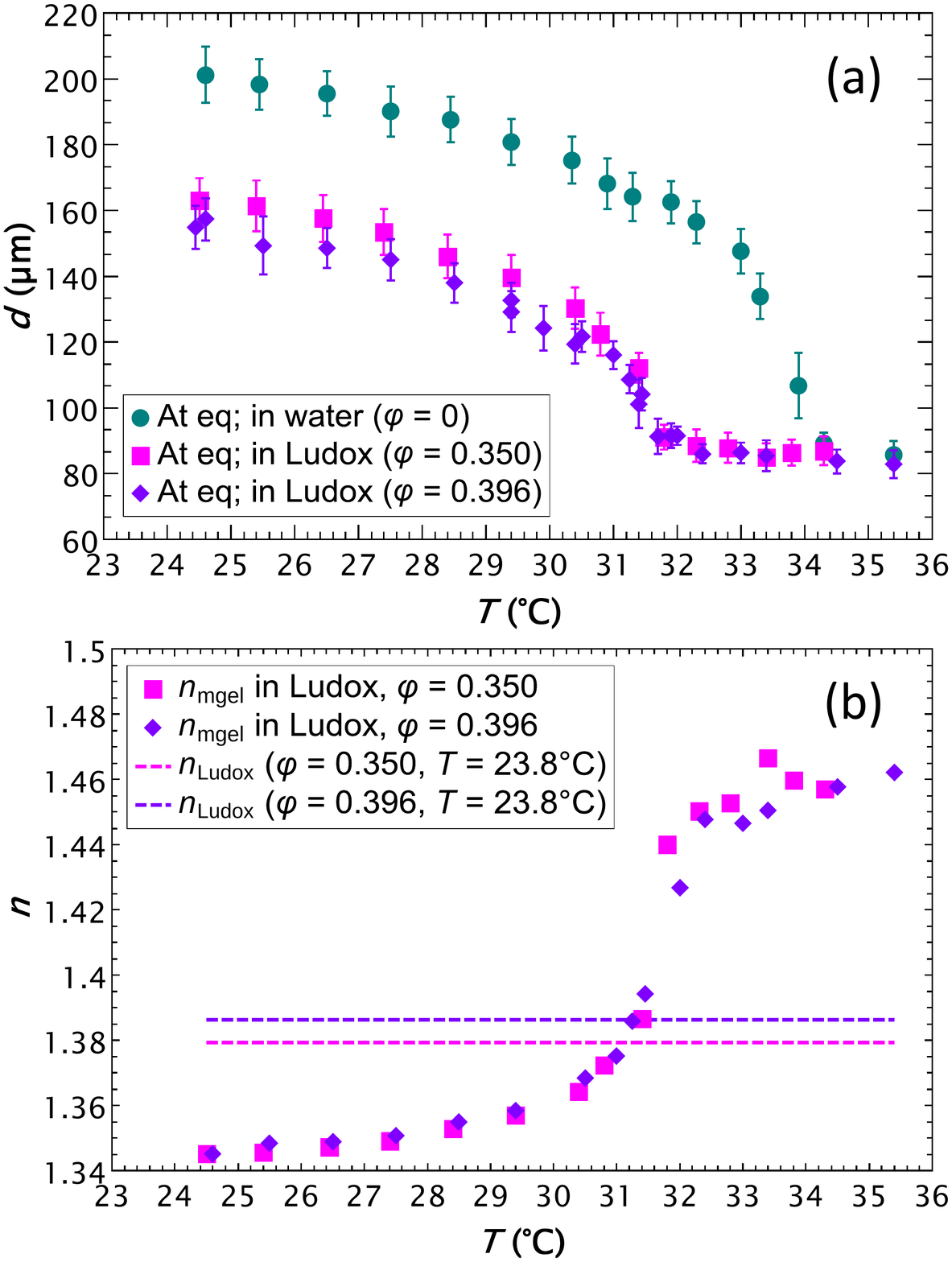}
\caption{\label{fig:D_n_vs_T} Measured size (a) and estimated refractive index $n_{\mathrm{mgel}}$ (b) of thermally-equilibrated PNiPAM mesogels as a function of the surrounding medium temperature in pure water or Ludox suspensions with $\varphi = 0.350$ and $\varphi = 0.396$. Error bars in (a) represent the standard deviation (SD). 
 In (b), the values of $n_{\mathrm{mgel}}$ (symbols) were estimated using Eq.~\ref{eq:refractive_index}. The dashed lines show the refractive index of the background Ludox suspensions.}
\end{figure}

We first show in Fig.~\ref{fig:ballsmicroscopy} the influence of temperature and the immersion medium on PNiPAM mesogels size and appearance. As expected,~\cite{kanai_preparation_2011} the size of the PNiPAM mesogels immersed in water decreases when the temperature increases. This behavior -- key to our approach -- is preserved when the mesogels are immersed in Ludox suspensions of various concentrations. In the latter, PNiPAM mesogels become less visible at temperatures around $31\mathrm{\degree C}$. The variation of PNiPAM mesogel volume indeed induces a change in their refractive index, which matches that of Ludox suspensions at temperatures around $31\mathrm{\degree C}$ (see discussion of Fig.~\ref{fig:D_n_vs_T}(b)).

PNiPAM mesogel size at equilibrium is plotted as a function of temperature and the immersion medium in Fig.~\ref{fig:D_n_vs_T}(a). For the three investigated suspension media, the size of the mesogels decreases when the temperature increases. In each case, the temperature at which the sharper change in size occurs -- usually called Volume Phase Transition Temperature (VPTT) -- was determined as the inflection point of $d(T)$, and the obtained values are shown in Table~\ref{tab:VPPT_Tcrossover_for_n}. The VPTT of our mesogels in water is  $33.6\mathrm{\degree C}$ and belongs to the range of $31.5-34\mathrm{\degree C}$ typically found for PNiPAM materials in water-only environments.~\cite{kanai_preparation_2011, lopez-leon_cationic_2006, senff_temperature_1999, kokufuta_effects_1993,okajima_kinetics_2002}

At low temperatures, the size of the mesogels is smaller in Ludox suspensions than in water, and the VPTT decreases as the Ludox concentration is increased. These differences in size and in VPTT are likely to be induced by the electrolytes present in the Ludox suspensions as a result of the charge stabilization performed by the manufacturer (we estimate from the data sheet that the concentration of the main electrolyte, $\mathrm{Na^+}$, is $ca.~0.1~\mathrm{M}$~\cite{Ludox_spec}). Previous studies have indeed shown that, upon salt addition, (i) the size of PNiPAM microgels decreases at a given temperature, and (ii) the volume phase transition is shifted towards lower temperatures.~\cite{lopez-leon_cationic_2006, park_sodium_1993, freitag_salt_2002} Our data are in qualitative agreement with these works. The main mechanism responsible for these changes is the dehydration of the PNiPAM chains by the added free ions, which leads to a smaller size at the lowest $T$ and promotes the volume transition at lower temperatures.~\cite{lopez-leon_cationic_2006, park_sodium_1993, freitag_salt_2002}

To investigate the close index matching of PNiPAM mesogels in Ludox suspensions around $31\mathrm{\degree C}$ shown in Fig.~\ref{fig:ballsmicroscopy}, their refractive index was estimated as a function of temperature ($T$) using

\begin{equation}
\label{eq:refractive_index}
n_{\mathrm{mgel}}(T)=x(T)\,\,n_{\mathrm{mono}} + [1-x(T)]  n_{\mathrm{w}}
\end{equation}
where $n_i$ is the refractive index of the object or fluid $i$ ($i$ = `mgel', `mono', and `w' for the mesogel, NiPAM monomer and water, respectively), and  $x$ is the volume fraction of NiPAM monomers within a mesogel sphere, calculated from the monomer concentration used in the synthesis as explained in Sec.~III of the SM. In Eq.~\ref{eq:refractive_index}, we use  $n_{\mathrm{mono}}=1.52$,~\cite{li_submicrometric_2015} and $n_{\mathrm{w}} = 1.325$, the latter measured at $T=23.8^{\circ}\mathrm{C}$. The variation of the refractive index of both NiPAM monomers and water is considered negligible across the investigated temperature range.

The so-estimated values of the PNiPAM mesogel refractive index are plotted as a function of temperature in Fig.~\ref{fig:D_n_vs_T}(b) (symbols), together with the refractive index values measured for the two Ludox suspensions (dashed lines). In both cases, $n_{\mathrm{mgel}}$ is lower than that of the surrounding medium at low temperatures, where the mesogels are highly swollen, such that $n_{\mathrm{mgel}}$ is close to the refractive index of water. Upon increasing the temperature, $n_{\mathrm{mgel}}$ increases and becomes higher than that of the surrounding medium. The crossover -- indicative of index matching conditions -- takes place at temperatures around $31\mathrm{\degree C}$ (values are provided in Table~\ref{tab:VPPT_Tcrossover_for_n}), corresponding to the temperature range where the mesogels almost `disappear' (see mid and bottom rows of Fig.~\ref{fig:ballsmicroscopy}). It is worth noting that, as the water refractive index is lower than that of the mesogels for all investigated temperatures, these conditions are never met in water and the balls never `disappear' when they are suspended in water (top row of Fig.~\ref{fig:ballsmicroscopy}). Finally, we emphasize that the fact that mesogels in Ludox suspensions meet index-matching conditions at intermediate temperatures  confirms that the silica particles cannot penetrate the mesogels. Indeed, if a significant amount of Ludox could penetrate the mesogels, $n_{\mathrm{mgel}}$ would be larger than the refractive index of the background suspension at all $T$ and index matching would never occur.

\begin{table}
\caption{\label{tab:VPPT_Tcrossover_for_n} Volume Phase Transition Temperature (VPTT) and temperature at which near index matching occurs for equilibrated PNiPAM mesogels dispersed in different media. }
\begin{ruledtabular}
\begin{tabular}{lcc}
Background medium&VPTT (°C)&$T_{\mathrm{index~matching}}$ (°C)\\
\hline
Water & $33.6 \pm 0.1$ & ---\\
Commercial Ludox ($\varphi=0.350$) & $31.4\pm$ 0.1 & $31.1\pm 0.2$\\
Concentrated Ludox ($\varphi=0.396$) & $31.3\pm$ 0.1 & $31.3\pm 0.2$\\
\end{tabular}
\end{ruledtabular}
\end{table}

Our final goal is to use PNiPAM mesogel ability to change size with temperature to access a wide range of Ludox volume fractions in a controlled manner using a single sample, as well as to impose well-controlled variations of the volume fraction with time (\textit{e.g.} quenches and ramps~\cite{peng_comparison_2014, peng_physical_2016, di_dynamics_2014} or cycles~\cite{tjhung_discontinuous_2017}). To this end, it is important to asses whether a unique, well-defined relationship between temperature and Ludox volume fraction holds, \textit{i.e.} to check whether the mesogel size at a given $T$ is reproducible and independent of the thermal history imposed to the sample.


\begin{figure} 
\includegraphics[width=\columnwidth]{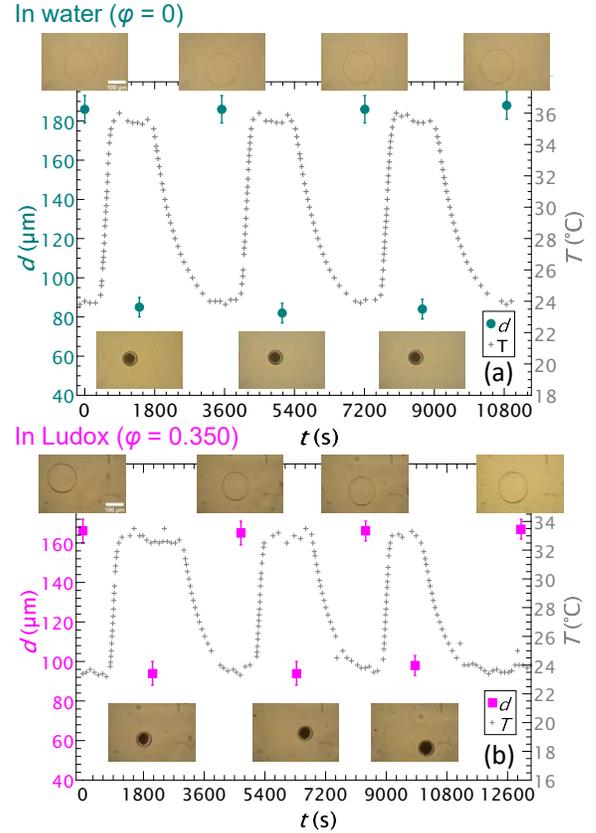}
\caption{\label{fig:T_cycles} Appearance and size of PNiPAM mesogels upon successive identical temperature cycles when suspended in water (a) or in a commercial Ludox suspension ($\varphi = 0.350$) (b). $\dot{T}_{\mathrm{up}}$ ranges from $2.3$ to $2.8\mathrm{\degree C/min}$, and $\dot{T}_{\mathrm{down}}$ from $-0.7$ to $-0.5\mathrm{\degree C/min}$ depending on the cycle. Error bars represent the SD. 
$t=0$ corresponds to the time at which micrographs were taken to assess mesogel size at equilibrium prior to $T$ cycles. Micrographs of a given mesogel are provided for each data point. Micrographs were all collected at magnification $\times 20$ and the scale bar (100 $\um$) applies to all the micrographs.}
\end{figure}

\begin{figure*} 
\includegraphics{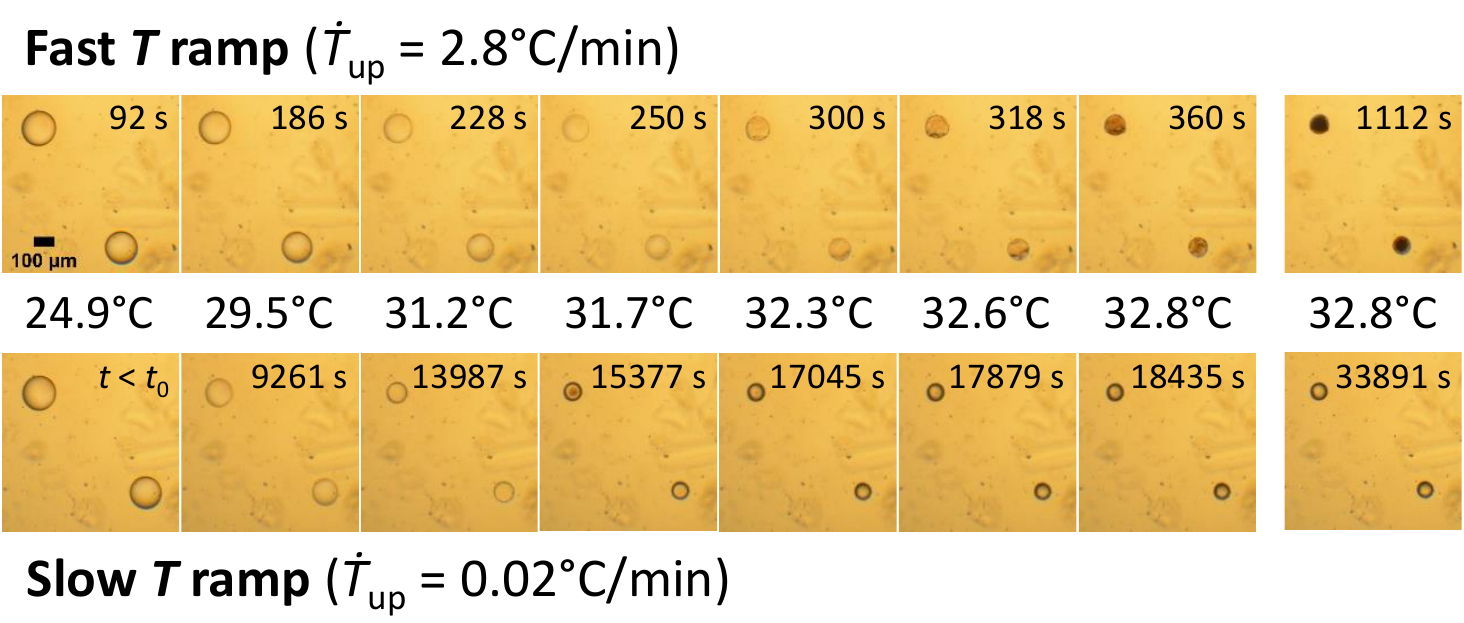}
\caption{\label{fig:fast_slow_cycle} Micrographs of two mesogels suspended in a commercial Ludox suspension ($\varphi = 0.350$) undergoing a fast $T$ ramp up (top part; $\dot{T}_{\mathrm{up}} = 2.8\mathrm{\degree C/min}$) and a slow $T$ ramp up (bottom part; $\dot{T}_{\mathrm{up}} = 0.02\mathrm{\degree C/min}$). Micrographs were collected at magnification $\times 3.5$. The scale bar applies to all of them.}
\end{figure*}

We show in Fig.~\ref{fig:T_cycles} the influence of successive identical temperature cycles on PNiPAM mesogel appearance and size in both water and the commercial Ludox suspension. In both cases, the mesogels are found to recover the same size and appearance after each $T$ ramp up and each $T$ ramp down. One may notice that the appearance of the mesogels during the high temperature plateau (bottom row of images in Fig.~\ref{fig:T_cycles}(a),(b)) is different than that seen in Fig.~\ref{fig:ballsmicroscopy}. This is due to the fact that the way the mesogels undergo the volume phase transition upon an increase in temperature depends on the rate $\dot{T}_{\mathrm{up}}$, as illustrated in Fig.~\ref{fig:fast_slow_cycle} for mesogels suspended in a commercial Ludox suspension (similar results are obtained for mesogels suspended in pure water -- data not shown). When the temperature is increased slowly (Fig.~\ref{fig:fast_slow_cycle}, bottom part), the mesogel appearance is very similar to that of thermally-equilibrated mesogels (see Fig.~\ref{fig:ballsmicroscopy}). Overall, the volume phase transition appears to be a smooth and continuous process. By contrast, when the temperature is increased quickly (Fig.~\ref{fig:fast_slow_cycle}, upper part), the change in mesogel size is very different. Once the mesogel temperature reaches that of the volume phase transition, its surface is deformed by the escaping water, forming some sorts of transient `bubbles'. At the end of the transition, the mesogels look dark, indicating that their structure is heterogeneous on length scales comparable to or larger than the wavelength of visible light. If the mesogels are kept at a high temperature, their appearance keeps on evolving (from the mesogel surface towards its centre) and finally becomes much lighter, identical to that of the mesogels which have been subjected to a slow increase in temperature (see Fig.~\ref{fig:ballsmicroscopy} and bottom part of Fig.~\ref{fig:fast_slow_cycle}), suggesting that rearrangements have occurred and led to a more homogeneous structure.

The formation of `bubbles' upon a quick increase in $T$ has been observed for many types of PNiPAM materials (\textit{e.g.} mesogels,~\cite{mou_monodisperse_2014, sato_matsuo_kinetics_1988} cylinders~\cite{okajima_kinetics_2002} and disks~\cite{kaneko_temperature-responsive_1995} with diameters of $100~\mathrm{\mu m}-1.0~\mathrm{mm}$,  $\sim1.1~\mathrm{mm}$ and $\sim15~\mathrm{mm}$, respectively) prepared with a small crosslinker/monomer ratio ($2\% \mathrm{mol.}$ in our case) and is associated with a three-stage shrinking process,~\cite{mou_monodisperse_2014, sato_matsuo_kinetics_1988} which can be seen in the video in Fig.~S5 (Multimedia view) in SM and is further discussed in Sec. IV of that same document. Our videos where very fast heating is carried out using a hair-dryer (Fig.~S6 (Multimedia views) in SM) show that the three-step process becomes more and more pronounced as $\dot{T}_{\mathrm{up}}$ increases. In the videos, water release at the third stage can be visualised due to both the high speed at which water is expelled out of the mesogels and the difference in refractive index between water and the Ludox suspension. %
%

\begin{figure} 
\includegraphics[width=\columnwidth]{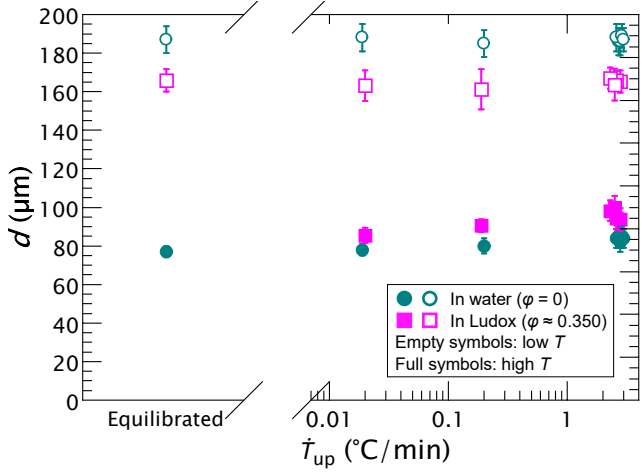}
\caption{\label{fig:T_rate} Mesogel size in water and in a commercial Ludox suspension ($\varphi = 0.350$) after $T$ ramps up (full symbols; high $T$\resub{, corresponding to $T = 35.4\pm 0.2~^{\circ}\mathrm{C}$ and $T = 33.0\pm 0.4~^{\circ}\mathrm{C}$ for water and the Ludox suspension as background fluids, respectively}) and down (empty symbols; low $T$\resub{, corresponding to $T = 23.8\pm 0.3~^{\circ}\mathrm{C}$ and $T = 23.6\pm 0.2~^{\circ}\mathrm{C}$ for water and the Ludox suspension as background fluids, respectively). Data are plotted }as a function of the rate at which the temperature is varied during the ramp up, $\dot{T}_{\mathrm{up}}$. Ramps down were all performed at $\dot{T}_{\mathrm{down}} \approx -0.5\mathrm{\degree C/min}$. Equilibration times prior to data collection for mesogel sizing were $\sim15~\mathrm{min}$ and $\sim10~\mathrm{min}$ at high and low $T$, respectively, except for the `Equilibrated' data. The latter, on the left of the axis break, correspond to data collected prior to any temperature cycle (empty symbols; low $T$) or after a $215$ min $T$ plateau (full symbol; high $T$). 
The set of points at the highest $\dot{T}$ also includes the data of Fig.~\ref{fig:T_cycles}, and error bars represent SD.}
\end{figure}

Although the appearance (and hence the microscopic structure) of the mesogels depends on the rate at which temperature is increased, Fig.~\ref{fig:T_rate} shows that their final size does not depend on that rate. Indeed, after $T$ ramps up with rates $\dot{T}_{\mathrm{up}}$ varied between $0.02$ to $3\mathrm{\degree C/min}$, the mesogel diameter reaches essentially the same value of $\sim80~\mathrm{\mu m}$ (solid symbols in Fig.~\ref{fig:T_rate}), independent of $\dot{T}_{\mathrm{up}}$, and equal to that obtained after the samples stayed at high $T$ for a couple of hours (`Equilibrated' data in Fig.~\ref{fig:T_rate}). After the samples have been cooled back to \resub{low $T$ (open symbols, see caption of Fig.~\ref{fig:T_rate})}, the mesogels recover their diameter of $\sim185~\mathrm{\mu m}$ and $\sim165~\mathrm{\mu m}$ in water and in the commercial Ludox suspension, respectively, similar to those they had prior to any $T$ cycle (`Equilibrated' data). These results, together with those presented in Fig.~\ref{fig:T_cycles}, confirm the ability of the mesogels to recover their size at either low or high $T$, regardless of the imposed thermal history.

\resub{Our microscopy observations suggest that, for fast enough ramps, the time scale of mesogel volume change is limited by the time required by the solvent to enter or leave the mesogel. By measuring the time evolution of $d$ while imposing upwards or downwards $T$ ramps between the fully swollen and fully shrunk states, we find that for $\dot{T}_{\mathrm{up}} \ge 2.7 \mathrm{\degree C/min}$ (resp., $|\dot{T}_{\mathrm{down}}| \ge 1.3 \mathrm{\degree C/min}$), the final size is reached up to 300 s (resp., up to 130 s) after attaining the target temperature. Note that, in the case of fast shrinking, a longer time scale of the order of 1.5 h is needed for the mesogels to recover their transparent appearance. By contrast, for slower upwards or downwards ramps, $d$ smoothly follows the evolution of $T$, with no further changes once the final $T$ is reached. }

Now that we have characterized the behavior of the mesogels in Ludox suspensions upon $T$ changes and demonstrated that \resub{the stationary state} is independent from the $T$ history, we investigate the dynamics of the Ludox nanoparticles in our mesogel-Ludox mixtures with DLS. We use mesogels from another batch than that used for optical microscopy, see Sec.~I.B in SM. Note that the dynamics we probe are those of the Ludox suspension rather than those of the mesogels. Indeed, our experiments are carried out at a scattering angle of $90 \degree$, where the mesogels do not scatter significantly due to their large size. Furthermore, our DLS set-up includes an imaging collection optics and a CMOS camera, allowing us to take space-resolved speckle pictures of the scattering volume over time, which are then processed to obtain the intensity auto-correlation (IAC) functions.~\cite{cipelletti_scattering_2016} When selecting the regions of interest (ROIs) to process these images, we make sure that they are free from mesogels at all times. An example of an image collected during DLS measurements showing the scattering volume, the mesogels and ROI selection is available in Fig.~S7 in SM.

\begin{figure} 
\includegraphics[width=\columnwidth]{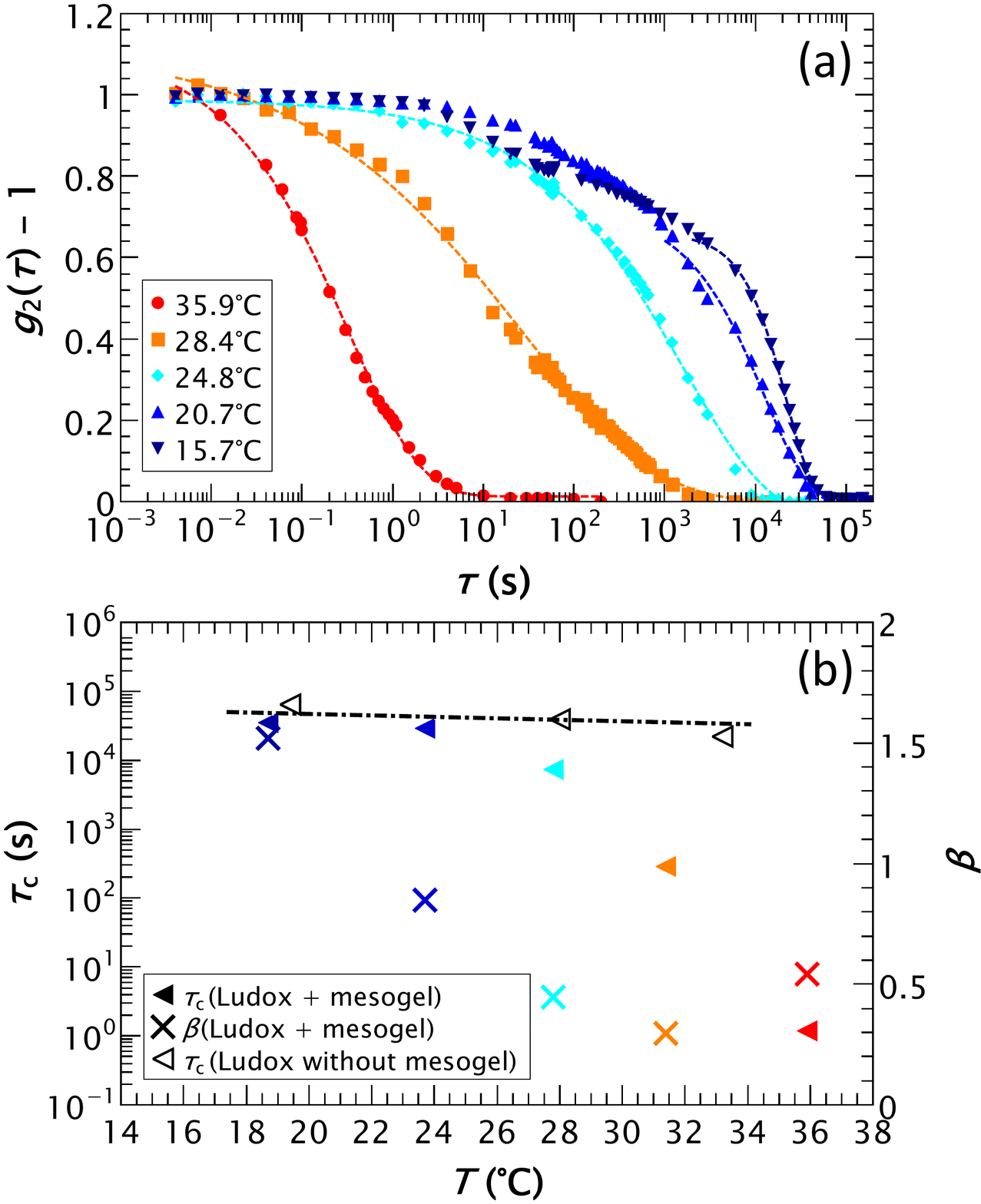}
\caption{\label{fig:DLS_old_sample} DLS data collected accross a wide range of temperatures on a concentrated Ludox suspension containing PNiPAM mesogels. Details about the mesogels used for this series of experiments are provided in the Sec.~I.B of the Supplementary Material. (a) Intensity auto-correlation (IAC) functions (symbols) and their fits (lines), obtained using Eq.~\ref{eq:fit_g2-1}. IACs functions were normalized to the smallest available delay time to bring their intercepts to unity.  (b) Fitting parameters $\tau_{\mathrm{c}}$ (full triangles) and $\beta$ (crosses) as a function of $T$ for the studied mesogels/Ludox mixtures. The color code is the same as in panel (a). The relaxation time $\tau_{\mathrm{c}}$ of a Ludox sample with no added mesogels ($\varphi=0.395$) measured at three different temperatures is also shown (empty triangles), together with the expected scaling with $T$ due to changes in thermal energy and solvent viscosity (dashed-dotted line).}
\end{figure}

Fig.~\ref{fig:DLS_old_sample}(a) shows the IAC functions collected for our mesogel-Ludox mixture. 
As $T$ decreases, the IAC curves decay at longer lag times, indicating slower dynamics of the Ludox particles. This behavior is consistent with what is expected from the $T$ dependence of the mesogels diameter: as temperature decreases, the mesogels become bigger and occupy more volume in the sample. Hence, the volume available to Ludox particles is smaller, and their effective volume fraction $\varphi$ increases, leading to slower dynamics~\cite{philippe_glass_2018}.

Interestingly, during the DLS measurements at the highest temperature ($T = 35.9 \mathrm{\degree C}$, above the VPTT), movies of the speckle images showed that PNiPAM mesogels moved throughout the sample (see Fig.~S8 (Multimedia views) in SM), which we explain by the relatively low background medium viscosity. Indeed, as discussed above, when $T$ increases, $\varphi$ decreases, which leads to a decrease in the viscosity of the Ludox suspension the mesogels are suspended in. Two types of motion were observed: creaming at first (Fig.~S8(a) in SM), followed by convection (Fig.~S8(b) in SM). Creaming is due to the density mismatch between the Ludox suspension and the PNiPAM mesogels; the latter having a lower density. Convection is likely to be due to slight local heating of the sample, due to the (small) absorption of laser light by PNiPAM. We find that convection only sets in after illuminating the sample for extended periods of time ($\sim 4~\mathrm{h}$) at the maximum laser power. To avoid convection, the laser power was decreased from $150~\mathrm{mW}$ to $37.5~\mathrm{mW}$ for all the measurements performed at lower temperatures, starting from $T = 28.4\mathrm{\degree C}$, and no creaming nor convection was observed over the time of the measurements. Importantly, even at the highest temperature, we find that mesogel motion is slow enough not to perturb the dynamics of the Ludox nanoparticles (see Fig.~S9 and Sec.~V.B in SM for details).

To quantify the $T$ dependence of the dynamics, we fit the decay of the IAC functions with
\begin{equation}
\label{eq:fit_g2-1}
g_{2}(\tau)-1=\left \{A \exp[-(\tau/\tau_{\mathrm{c}})^{\beta}]\right \}^{2}
\end{equation}
where $A$ is the amplitude of the relaxation mode, $\tau_{\mathrm{c}}$ its relaxation time and $\beta$ the stretching exponent. Fits using Eq.~\ref{eq:fit_g2-1} are shown as dashed lines in Fig.~\ref{fig:DLS_old_sample}(a).
For IAC data collected at the lowest temperatures ($T \le 20.7\mathrm{\degree C}$), two relaxation modes were clearly observed, a distinctive feature of the slow dynamics of supercooled systems.~\cite{biroli_perspective_2013,hunter_physics_2012} In this case, we fitted the slowest relaxation mode. The fitting parameters $\tau_{\mathrm{c}}$ and $\beta$ are shown in Fig.~\ref{fig:DLS_old_sample}(b). Across the investigated $T$ range, $\tau_{\mathrm{c}}$ increases as $T$ decreases, spanning more than four decades. Note that, for all $T$, $\tau_{\mathrm{c}}$ is significantly larger than 0.18 ms, the relaxation time obtained from Eq.~\ref{eq:fit_g2-1} in the infinite dilute limit. To confirm that the wide variation of $\tau_{\mathrm{c}}$ is due to the variation of the mesogel size, values of $\tau_{\mathrm{c}}$ measured at different $T$ on a concentrated Ludox suspension  without added mesogels are also shown ($\varphi=0.395$, open symbols in Fig.~\ref{fig:DLS_old_sample}(b)). As $T$ decreases, $\tau_{\mathrm{c}}$ increases only slightly, indicating a marginal slowing down of the dynamics when $T$ is lowered. Note that this variation is orders of magnitude smaller than that observed in the Ludox sample containing the mesogels. Indeed, in the case of the pure Ludox sample, $\tau_{\mathrm{c}}$ is multiplied by $\sim2.8$ when $T$ decreases from $33.2\mathrm{\degree C}$ to $19.4\mathrm{\degree C}$, while it is multiplied by more than $800$ over the same $T$ range in the case of the mesogels/Ludox mixture. At fixed $\varphi$, the relaxation time of a colloidal suspension is expected to scale as $D_0^{-1} \propto \eta_0/T$, with $D_0$ the infinite dilution particle diffusion coefficient and $\eta_0$ the ($T$-dependent) solvent viscosity. The dashed-dotted line in Fig.~\ref{fig:DLS_old_sample}(b) shows the variation of $\eta_0/T$ over the measured $T$ range: it accounts for most of the observed change of $\tau_{\mathrm{c}}$ in the Ludox suspension without mesogels, confirming that the dramatic change in relaxation time of the Ludox-mesogels mixtures is due to the volume change of the mesogels upon varying temperature.


As a further proof that our mesogel-Ludox mixture exhibits a broad range of dynamical behaviors with $T$, we look into the $T$-dependence of the stretching exponent $\beta$. When $T$ decreases (\textit{i.e.} $\varphi$ increases), $\beta$ first decreases, reaching a value as low as $\sim0.3$ at $T=31.4\mathrm{\degree C}$, and then increases again, reaching a value of $1.5$ at $T=18.7\mathrm{\degree C}$. These variations in $\beta$ are similar to those reported by Philippe \textit{et al.}~\cite{philippe_glass_2018} for a pure Ludox suspension, where the volume fraction was varied by preparing distinct samples. Values of $\beta$ below 1 -- characteristic of a stretched exponential relaxation -- are a feature of the intermediate $\varphi$ regime observed in Ref.~\cite{philippe_glass_2018} (termed `regime II' therein), which corresponds to the supercooled regime and where $\tau_{\mathrm{c}}$ increases sharply with volume fraction. By contrast, values of $\beta$ above 1 -- characteristic of a compressed exponential relaxation -- are a feature of high-$\varphi$ regime (`regime III' in Ref.~\cite{philippe_glass_2018}), where all the samples are in a glassy state and $\tau_{\mathrm{c}}$ is nearly independent of the volume fraction. These results fully demonstrate that varying $T$ over a few degrees in a single Ludox-mesogels mixture allows us to access the same states as those obtained for pure Ludox suspensions prepared at different volume fractions.

Using $\tau_{c} = f(\varphi)$ data collected by Philippe \textit{et al.}~\cite{philippe_glass_2018} for pure Ludox suspensions and the values of $\tau_{c}$ of our mesogel-Ludox sample, we estimate $\Delta \varphi$, the maximum variation in the Ludox volume fraction \resub{corresponding} to the range of $\tau_{c}$ measured \resub{when varying $T$ in the Ludox-mesogel mixture of Fig.~\ref{fig:DLS_old_sample}}. We find $\Delta \varphi \approx 3.5\%$ of the same order of magnitude but somehow larger than $\Delta \varphi \approx 1.2\%$, the value estimated from the volumes and volume fractions of the stock mesogel and the stock Ludox suspensions used to prepare the sample (see details in Sec.~VI in SM). Three hypotheses may explain this discrepancy: (i) The estimation of $\Delta \varphi$ is based on the assumption that the behavior of our Ludox sample is exactly the same as that studied by Philippe \textit{et al.},~\cite{philippe_glass_2018} which may not be the case. Indeed, mesogel addition is accompanied by a small dilution of the Ludox sample due to the simultaneous addition of a small amount of water, which may affect electrolyte concentrations and thus electrostatic screening and Ludox surface charge. Furthermore, the experiments presented here were performed using a batch of Ludox particles different from that of Ref.~\cite{philippe_glass_2018}, and commercial suspensions are known to exhibit batch-to-batch differences;  (ii) As previously discussed, creaming has occurred in the sample at the highest investigated temperature, which is also the first temperature the sample was subjected to. This resulted in a greater mesogel concentration near the top of the sample, where the scattering volume is located, as compared to the average mesogel concentration used to estimate $\Delta \varphi$; 
(iii) Various hypotheses and approximations (see Sec.~VI in SM for details) are required to estimate both values of $\Delta \varphi$, which probably also contribute to the observed difference.

\section{\label{sec:conclusion}Conclusions and perspectives}

We have successfully prepared a single suspension of silica nanoparticles that exhibits a broad range of dynamical behaviors upon varying $T$, showing characteristics similar to those obtained for a series of distinct Ludox-only samples with $\varphi \approx 0.367-0.403$. This was achieved by adding a relatively small amount of PNiPAM mesogels ($\sim 2-5\%~$vol., depending on $T$) 
to a concentrated colloidal suspension. While the same result could likely be achieved by immersing a macroscopic piece of PNiPAM gel in the sample, our $\sim 200 \um$ mesogels offer several advantages. First, they allow for a faster change of volume of PNiPAM, since the swelling time of a gel in a good solvent is proportional to the square of its linear size.~\cite{tanaka_kinetics_1979} Second, any change in PNiPAM volume creates a local gradient of nanoparticle concentration, which will relax slowly in dense suspensions. Mesogels allow for splitting the overall volume over which gradients occur in many smaller regions, thereby accelerating sample equilibration. Finally, mesogels/Ludox mixtures can easily be transferred in cells of arbitrary size and shape, \textit{e.g.} thin capillaries used for X-ray scattering, and may be used for rheology experiments, since the mesogel size is smaller than the gap of typical plate-plate or couette geometries.

\resub{The achievable change in volume fraction $\Delta \varphi$ is limited essentially by the amount of mesogels that one is willing to add to the sample. Absolute volume fraction changes larger than 20\% are in principle possible (see Fig. S9 in the SM). However, this would typically come at the expense of having 10\% or more of the total sample volume occupied by mesogels. Another factor limiting $\Delta \varphi$ is  the extent of the volume variation of the mesogels. We have shown that the swelling of mesogels in Ludox suspensions is reduced as compared to that in water. This is likely due to the electrolytes present in Ludox suspensions, although the osmotic pressure exerted by the colloids themselves may also play a role.
Finally, since $\varphi$ is tuned by varying temperature, a potential concern in experiments is the impact of (unwanted) temperature fluctuations. In Sec. VI of the SM, we show that for typical experimental conditions and considering a relatively large temperature fluctuation $\delta T = 0.1~^{\circ}\mathrm{C}$, the resulting change in colloid volume fraction is modest: of order $10^{-3}$ close to the VPTT and significantly smaller at the lower and higher ends of the typical $T$ range.}

A few points could still be improved to optimize our method, essentially concerning the control of the amount and distribution of mesogels in the sample. As discussed above, a poorly known amount of water is added to the nanoparticle suspension together with the mesogels. As a result, the amount of added mesogels is not known precisely, nor is the actual volume fraction of the nanoparticles. A solution that we are currently investigating consists in adding the mesogels in a freeze-dryed state and let them rehydrate in the final sample. This would have the twofold advantage of knowing precisely the amount of added mesogels (\textit{e.g.} by weighting and counting them) and avoiding any dilution of the nanoparticle suspension. Another strategy involves labelling the mesogels with a fluorophore,~\cite{kim_fluorescent_2015,conley_superresolution_2016} allowing for their visualization in the scattering volume. This approach would also address the issue of the uneven spatial distribution of mesogels due to creaming, since their number concentration would be directly determined in the very sample region where the nanoparticle dynamics are probed. The effect of creaming could also be mitigated by using a light scattering cell with a smaller height.

Although we have tested the method described here only on Ludox samples, we expect it to apply quite generally to any water-based colloidal suspensions, provided that the physico-chemistry of the system does not severely interferes with the swelling/deswelling capability of PNiPAM. The present work, where PNiPAM thermosensitivity was shown to still hold in the presence of electrolytes and at a basic pH $\sim 9$, together with previous works,~\cite{annakaSaltinducedVolumePhase2000,peiEffectPHLCST2004} suggest that PNiPAM swelling/deswelling is indeed preserved for the solvent conditions encountered in a wide range of suspensions.

We believe that the versatility of our method opens several research paths. Among various possibilities, we mention the question of how far a local disturbance of volume fraction (due to swelling/deswelling of a mesogel) has an impact on the dynamics of the surrounding suspension, which would be an experimentally new way to study the role of spatial correlations of the dynamics.~\cite{biroli_perspective_2013,li_anatomy_2020} By coupling rheology and DLS measurements,~\cite{pommella_coupling_2019} the method presented here could also allow assessing whether or not rejuvenation upon shear or $\varphi$ jumps are equivalent to each other in concentrated suspensions of soft nanoparticles, when avoiding complications potentially arising from particle interpenetration. Finally, the enhanced ease of handling of nanoparticle suspension at high $T$, due to the decrease in effective volume fraction, could allow for investigating the behavior of samples comprising particles smaller than those typically used so far. This should provide interesting insights about aging, as colloidal glasses made of smaller particles age quicker than those made of larger particles and should therefore lead to samples closer to equilibrium.~\cite{hunter_physics_2012, hallett_local_2018} Additionally, this would allow testing experimentally recent numerical findings that have unveiled intriguing differences in the non-linear mechanical properties of glasses depending on equilibration.~\cite{singh_brittle_2020}

\section*{Supplementary Material}
See supplementary material for details about PNiPAM mesogel production for both optical microscopy and DLS, micrographs of equilibrated PNiPAM mesogels at different temperatures and immersion media, `bubble' appearance during PNiPAM mesogel shrinkage following a quick increase in $T$, DLS data analysis (where the selection of regions of interest for data processing is discussed, together with creaming and convection at $T=35.9\mathrm{\degree C}$), and calculation of the effective volume fraction of the Ludox particles in the presence of mesogels. Video files showing showing mesogels suspended in Ludox suspensions undergoing shrinkage for $\dot{T}_{\mathrm{up}}=3 \mathrm{\degree C/min}$ at magnification $\times 20$, and upon heating up with a hair dryer at magnifications $\times 3.5$, $\times 10$ and $\times 20$ are also available, as well as video files showing creaming and convection observed during DLS measurements at $T=35.9\mathrm{\degree C}$.

\begin{acknowledgments}
This work was financially supported by CNES and L2C. L. Cipelletti gratefully acknowledges support from the Institut Universitaire de France. DOWSIL\textsuperscript{TM} RSN-0749 resin was kindly provided free-of-charge by Univar Solutions SAS. We are grateful to T. Phou and E. Chauveau for their assistance regarding PNiPAM mesogel synthesis, and to C. Blanc for his help with optical microscopy.

\end{acknowledgments}

\section*{Data Availability Statement}
The data that support the findings of this study are openly available in Zenodo at http://doi.org/10.5281/zenodo.5891491.

\newpage

\section{Supplementary Material to\\
Controlling the volume fraction of glass-forming colloidal suspensions using thermosensitive host `mesogels'}

\section{\label{sec:ball_prod}PNiPAM mesogel production}

\subsection{\label{sec:ball_prod_microscopy}Samples for optical microscopy}

Aqueous drops containing \textit{N}-isopropylacrylamide (NiPAM) monomers, a crosslinker and a photoinitiator were produced in silicone oil and stabilized with a surfactant. The drops were then polymerized under UV-light to form PNiPAM mesogels. To prepare the oil phase, $0.742~\mathrm{g}$ silicone surfactant DOWSIL\textsuperscript{TM} RSN-0749 resin ($3~wt\%)$ was added to $24~\mathrm{g}$ silicone oil 47 V 100. The oil phase was degassed with argon for $\sim 4~\mathrm{h}$. To prepare the aqueous solution, $6.2~\mathrm{mg}$ \textit{N,N'}-methylenebisacrylamide (BIS) crosslinker ($0.02~\mathrm{M}$) and $226~\mathrm{mg}$ NiPAM monomers ($1~\mathrm{M}$) were weighted and transferred in a brown glass vial (to avoid light-induced polymerization), prior to adding $6.2~\mathrm{\mu L}$ 2-hydroxy-2-methylpropiophenone photoniator ($0.3~wt\%)$ and $2.0~\mathrm{mL}$ deionized (DI) water. Argon was bubbled in the solution for $\sim 15~\mathrm{min}$ just before starting the synthesis.

\begin{figure*}[h!] 
\includegraphics{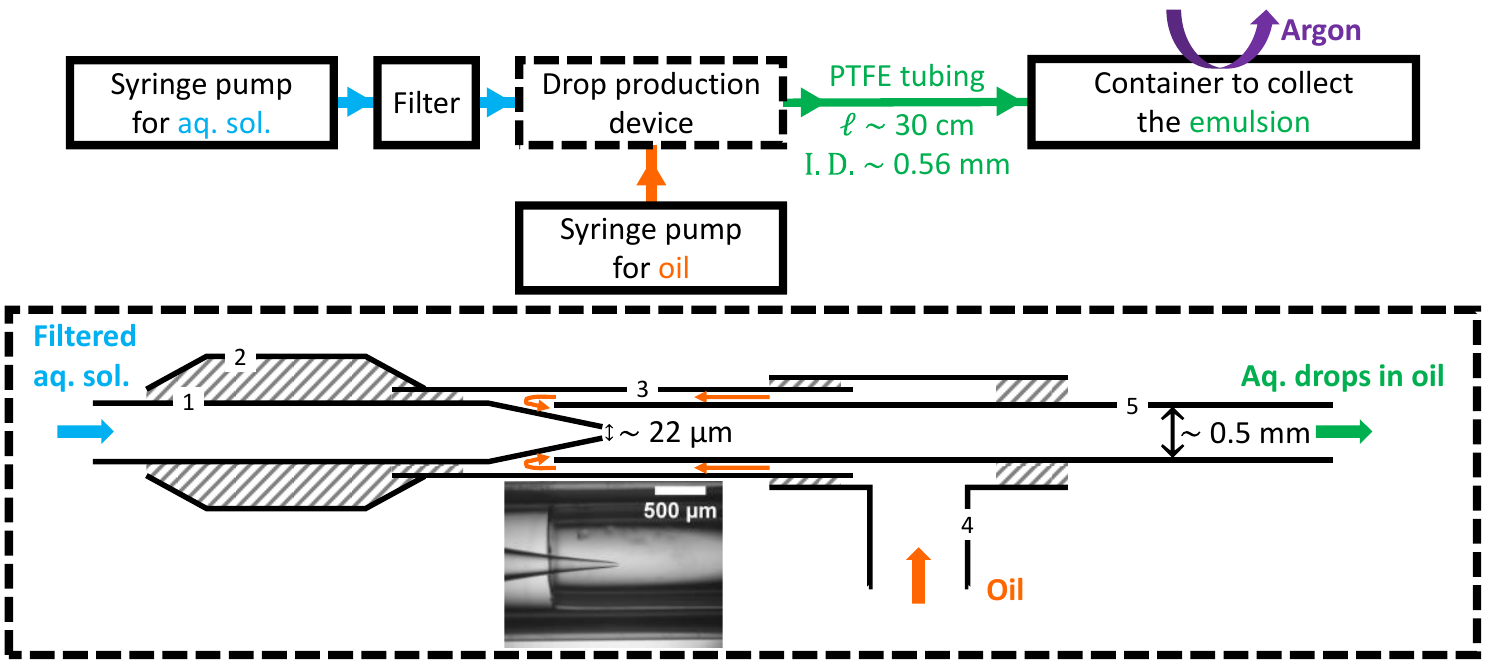}
\figuretag{S1}
\caption{\label{fig:setup} Schematic of the microfluidic setup for drop production. The bottom frame shows details about the home-made device where the drops are formed (dashed box in the upper part of the figure). 1: Stretched glass capillary with a tip diameter of $22\um$ (unstretched capillary ref: \textit{cf 5} hereafter); 2: Barbed connector, I.D: 3/32”, Harvard Apparatus (ref: 72-1426); 3: Glass capillary, O.D: $1.5-1.6~\mathrm{mm}$, I.D: $1.1-1.2~\mathrm{mm}$, Labbox (ref: MPC3-090-500); 4: T-connector, I.D: 1/8”, Harvard Apparatus (ref: 72-9282); 5: Glass capillary, O.D: $1.0~\mathrm{mm}$, I.D: $0.5~\mathrm{mm}$, l: $10~\mathrm{cm}$, World Precision Instruments (ref: TW100-4).}
\end{figure*}

The microfluidic setup used to produce the water-in-silicone oil emulsion is sketched in Fig.~\ref{fig:setup}. A $20~\mathrm{mL}$ plastic syringe (Omnifix\textsuperscript{\textregistered} Luer-Lock $20~\mathrm{mL}$; ref: T550.1; B. Braun) and a $2~\mathrm{mL}$ plastic syringe (Omnifix\textsuperscript{\textregistered} Luer-Lock $2~\mathrm{mL}$; ref: LY21.1 B. Braun) were loaded with the oil and the aqueous phases, respectively, connected to a home-built microfluid setup (see schematic at the bottom of Fig.~\ref{fig:setup}) and placed on two distinct syringe pumps (PHD2000 Infusion; ref: 70-2000; Harvard Apparatus). A $0.45~\mathrm{\mu m}$ pore size mixed cellulose ester syringe filter was inserted between the aqueous phase syringe and the corresponding entrance of the microfluidic device to prevent clogging. The flow rates on the syringe pumps were set to $10~\mathrm{mL/h}$ (syringe diameter: $19~\mathrm{mm}$) and $0.3~\mathrm{mL/h}$ (syringe diameter: $7~\mathrm{mm}$) for the oil and the aqueous phases, respectively. Once the steady-state was reached, the emulsion was collected in a container flushed with argon. Still under argon flush, the cap was removed and replaced by cling film prior to placing the container under a UV lamp (Dual Wave UV Analysis Lamp, 2 x 4 W, 254 nm and 365 nm; ref: H466.1; Herolab) for $\sim 1~\mathrm{h}$ for polymerization of the aqueous drops to take place, yielding mesogels. The cap was put back (still under argon) on the vial and the sample was left to rest overnight. The mesogels were then washed and transfered in water as described in Section II B 1 of the main text.

The room air conditioning temperature was set to $20\mathrm{\degree C}$ at all times during the synthesis, since the synthesis temperature influences the structure and appearance of PNiPAM-based materials. Previous research has shown that the structure of PNiPAM macrogels synthesized at temperatures below $25\mathrm{\degree C}$ is homogeneous on length scales of the order of the light wavelength and the materials appear relatively transparent in water, while the structure of PNiPAM macrogels synthesized at higher temperatures is heterogeneous upon that same scale and the materials are white.~\cite{hirokawa_direct_1999, hirokawa_sponge-like_2008, kayaman_structure_1998} We performed a synthesis at $30\mathrm{\degree C}$ to verify that this is also true for mesogels and indeed obtained mesogels that looked white in water at room temperature. Once suspended in Ludox, mesogels synthesized at $T=30\mathrm{\degree C}$ remained visible at all temperatures, contrary to the mesogels used in our study (see Figs~1 and 2(b) as well as related discussions in the main text). Finally it is worth mentioning that, assuming a homogeneous distribution of crosslinker within the mesogels, the average distance between two neighboring crosslinkers can be computed and it is equal to about 8 nm. For such a reason we expect a scarce Ludox nanoparticle diffusion, if any, into the mesogels.

\subsection{\label{sec:ball_prod_DLS}Samples for Dynamic Light Scattering (DLS)}

The synthesis of mesogels for the DLS study was similar to that described above, but polymerization was performed directly in a home-made microfluidic device similar to that shown in Fig.~\ref{fig:setup}, using a protocol adapted from that of Chen \textit{et al}.~\cite{CHEN2010168} A redox activator -- both oil- and water-soluble -- was added to the silicone oil, and subsequently diffused into the aqueous phase where it activated the initiator (see next paragraph for the composition of the two phases). The tip of the inner capillary where drops form was $18~\mathrm{\mu m}$ wide and the collecting tubing was $3.6~\mathrm{m}$ long (\textit{i.e.} significantly longer than for the synthesis with the photoinitiator described in Section~\ref{sec:ball_prod_microscopy}), to allow the aqueous drops to polymerize while travelling in the tubing before reaching the collecting container. The container was left to rest overnight prior to adding $1~\mathrm{mL}$ DI water, and washing the mesogels with diethyl ether, followed by DI water. No surfactant was used and no exposure to UV light was needed. The oil and aqueous phase flow rates were $7~\mathrm{mL/h}$ and $1~\mathrm{mL/h}$, respectively. Air conditioning was set to $24\mathrm{\degree C}$.

For this synthesis, the oil phase was prepared by dispersing $1.5~\mathrm{mL}$ tetramethylethylenediamine (TEMED) in $28.5~\mathrm{mL}$ silicone oil. The aqueous phase was prepared by dispersing $15.5~\mathrm{mg}$ \textit{N,N'}-methylenebisacrylamide (BIS) crosslinker ($0.05~\mathrm{M}$), $8.6~\mathrm{mg}$ potassium persulfate (KPS) initiator ($0.02~\mathrm{M}$) and $226~\mathrm{mg}$ NiPAM monomers in $2.0~\mathrm{mL}$ DI water.

The so-produced mesogels were spherical and looked relatively transparent in water. The mesogel/background medium interface was slightly less well defined than that of the mesogels prepared with the photoinitiated synthesis, which we attribute to the absence of surfactant. In water, the so-produced PNiPAM mesogels had a diameter of $221~\mathrm{\mu m}$ (SD: $11~\mathrm{\mu m}$) at room temperature, and in commercial Ludox ($\varphi = 0.350$), the diameters were $209~\mathrm{\mu m}$ (SD: $13~\mathrm{\mu m}$) and $149~\mathrm{\mu m}$ (SD: $10~\mathrm{\mu m}$) at room and high temperatures, respectively. The fact that the ratio between the mesogel diameter at high and low temperature, $d(T_{\mathrm{high}})/d(T_{\mathrm{low}})$, is higher in the present case as compared to that obtained for the mesogels prepared with the photoinitiated synthesis is due to the difference in the crosslinker/monomer ratios ($5~\mathrm{mol}\%$ and $2~\mathrm{mol}\%$, respectively), consistent with literature data.~\cite{okajima_kinetics_2002}

\begin{figure*}[h!] 
\includegraphics{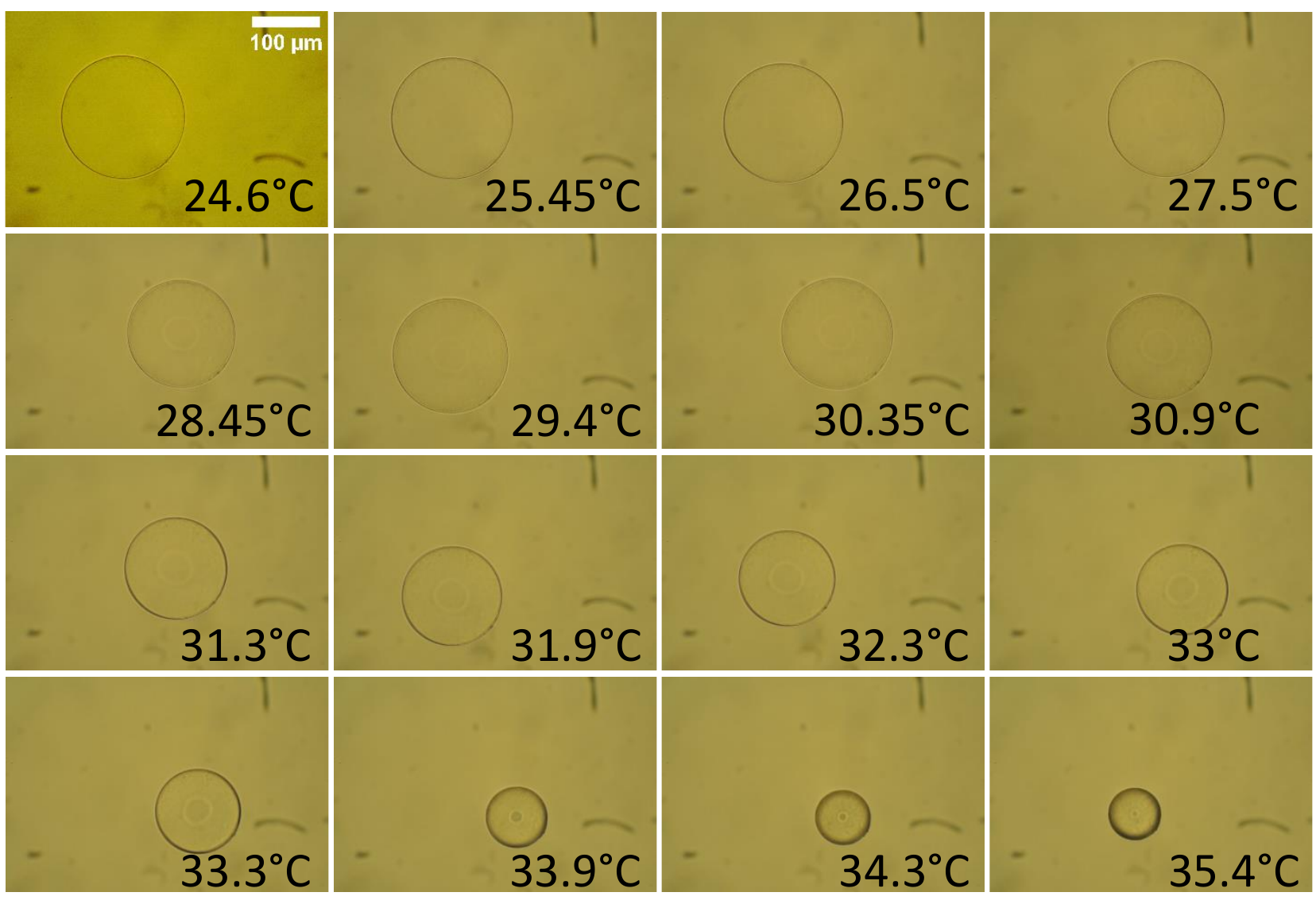}
\figuretag{S2}
\caption{\label{fig:balls_in_water} Micrographs of an equilibrated PNiPAM mesogel dispersed in water taken at different temperatures. Images were collected at magnification $\times20$. The scale bar applies to all of them.}
\end{figure*}

\section{\label{sec:extended_microscopy}Micrographs of equilibrated PNiPAM mesogels at different temperatures and immersion media}

\begin{figure*}[h!] 
\includegraphics{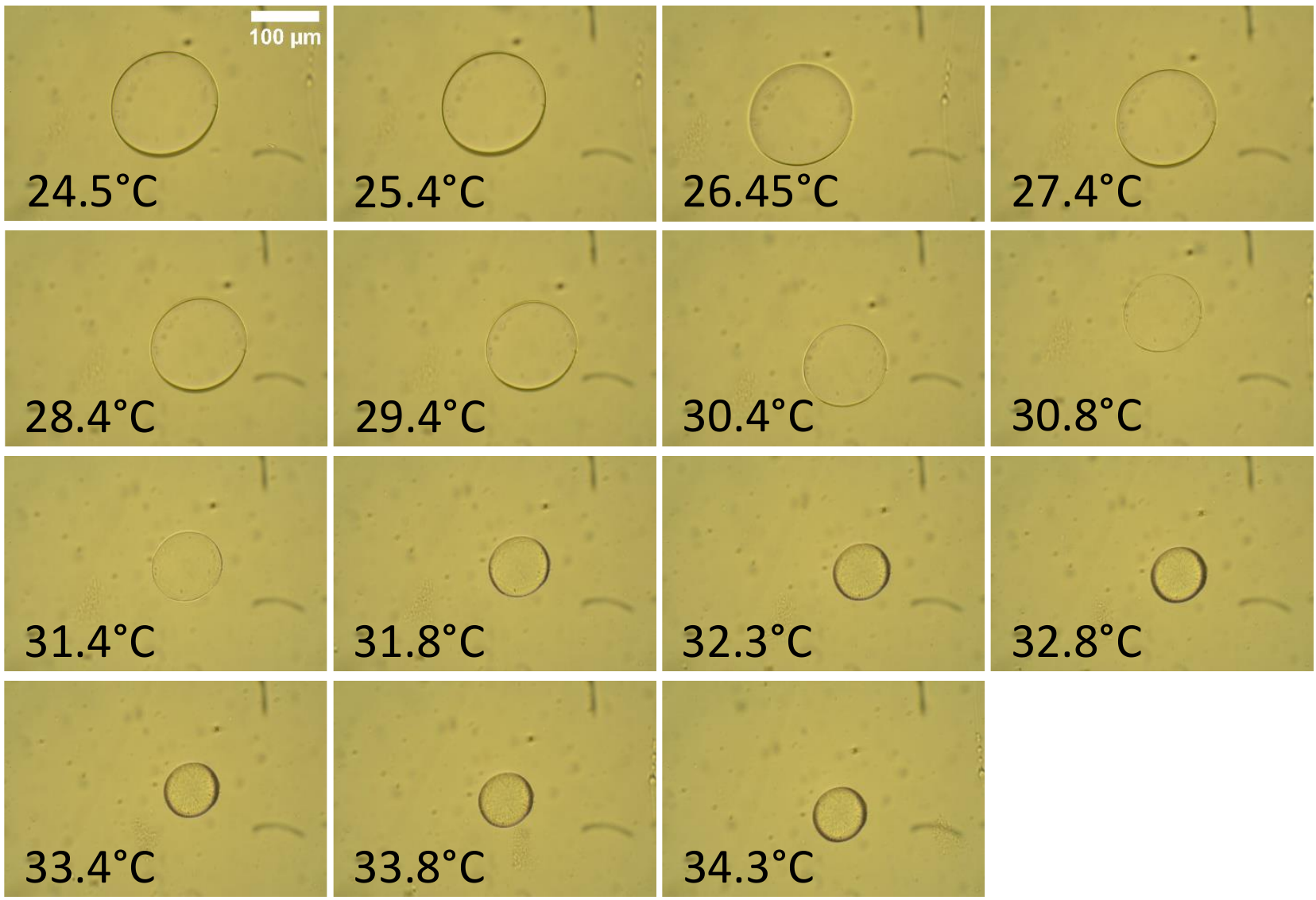}
\figuretag{S3}
\caption{\label{fig:balls_in_com_Ludox} Micrographs of an equilibrated PNiPAM mesogel dispersed in a commercial Ludox suspension ($\varphi = 0.350$) taken at different temperatures. Images were collected at magnification $\times20$. The scale bar applies to all of them.}
\end{figure*}

\begin{figure*}[h!] 
\includegraphics{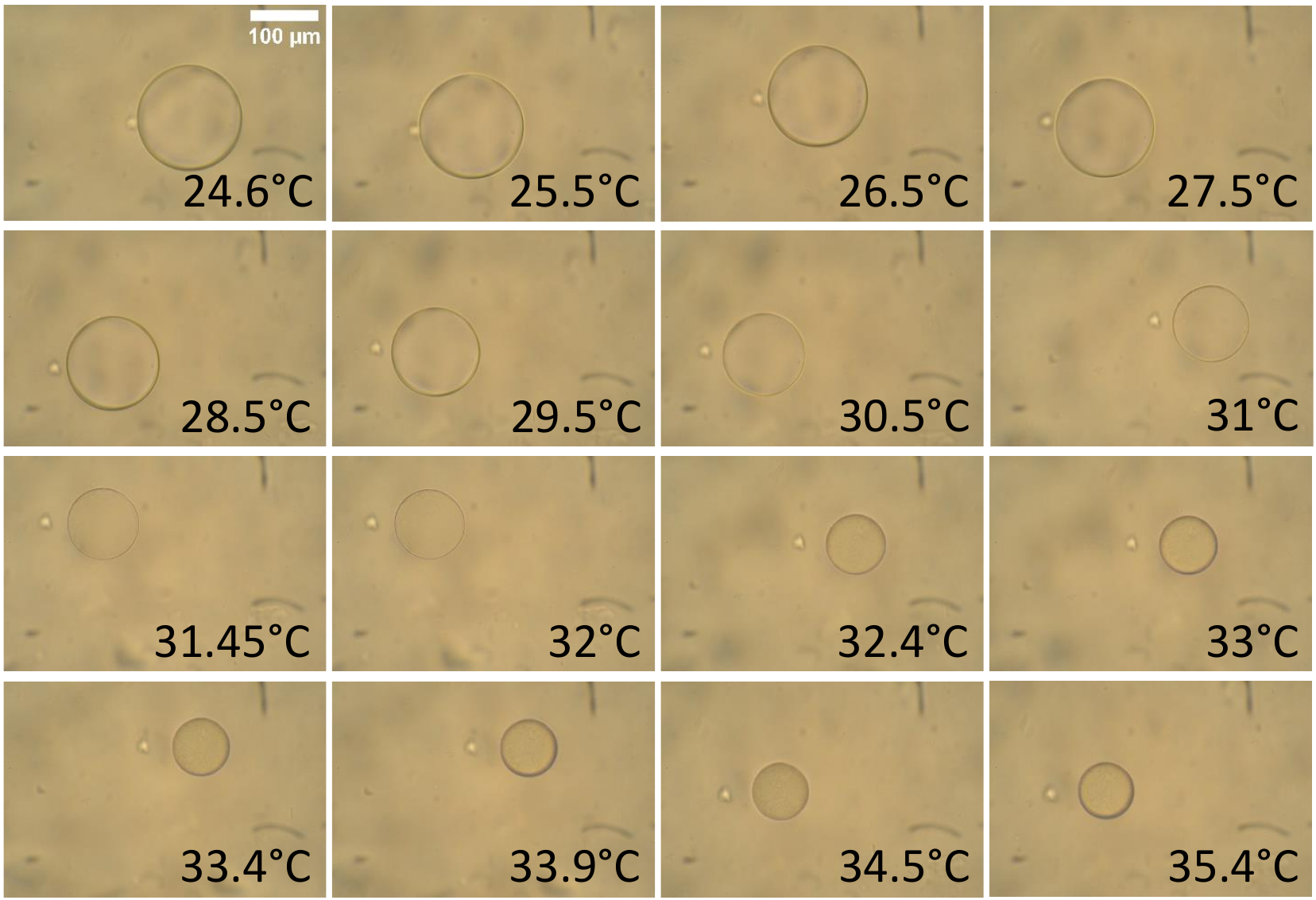}
\figuretag{S4}
\caption{\label{fig:balls_in_conc_Ludox} Micrographs of an equilibrated PNiPAM mesogel dispersed in a concentrated Ludox suspension ($\varphi = 0.396$) taken at different temperatures. Images were collected at magnification $\times20$. The scale bar applies to all of them.}
\end{figure*}

Micrographs of equilibrated PNiPAM mesogels synthesized with the protocol of Section~\ref{sec:ball_prod_microscopy}, taken at various temperatures and for different immersion media are shown in Figs.~\ref{fig:balls_in_water}-\ref{fig:balls_in_conc_Ludox}. These images complement Fig. 1 of the main text.


\section{\label{sec:calc_x_monol}Volume fraction $x$ of NiPAM monomers in a mesogel sphere}
To calculate the volume fraction $x$ of NiPAM monomers in a mesogel sphere introduced in Eq.~1 of the main text, we use the following expression:
\begin{equation}
\label{eq:x_mono}
x(T)=\frac{\frac{m_{\mathrm{NiPAM}}}{\rho_{\mathrm{NiPAM}}}}{\frac{\pi}{6}\pi[d(T)]^{3}} \,.
\tag{S1}
\end{equation}
In Eq.~\ref{eq:x_mono}, the numerator corresponds to the volume occupied by NiPAM monomers in a mesogel. It is obtained from the mass density of NiPAM,~\cite{sbeih_structural_2019} $\rho_{\mathrm{NiPAM}}=1.1~\mathrm{g.cm^{-3}}$, taken to be constant across the investigated temperature range, and the mass of NiPAM initially contained in each drop produced by the microfluidic device, $m_{\mathrm{NiPAM}}=c_{\mathrm{NiPAM}} V_{\mathrm{drop}}$, with $c_{\mathrm{NiPAM}}$ the concentration of NiPAM monomers in the aqueous solution injected in the microfluidic device and $V_{\mathrm{drop}}$ the drop volume. The denominator corresponds to the volume of a mesogel at temperature $T$, with $d(T)$ the diameter of the mesogel at that temperature (see Fig.~2(a) of the main paper).
\newpage
\section{\label{sec:bubbles}`Bubble' appearance during PNiPAM mesogel shrinkage following a quick increase in $T$}

\begin{figure*}[h!] 
\includegraphics{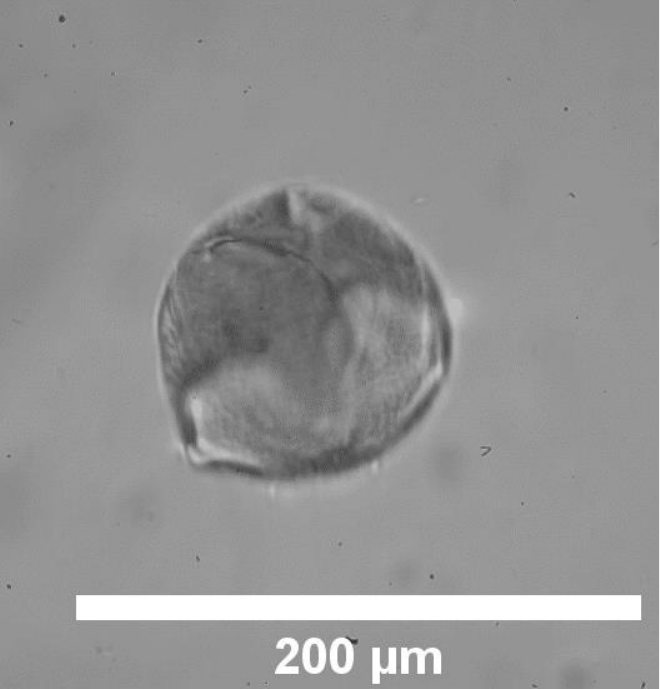}
\figuretag{S5}
\caption{\label{fig:Shrinking_3cpmin_x20} Frame taken from movie `FigS5\_Mesogel\_shrinking\_3Cpmin\_x20.mp4' -- available as an independent supplementary material file -- showing the shrinkage of a PNiPAM mesogel suspended in a commercial Ludox suspension ($\varphi=0.350$) and heated at $\dot{T}_{\mathrm{up}}=3\mathrm{\degree C/min}$. Magnification: $\times 20$. Field of view dimensions: $234~\mathrm{\mu m} \times 244~\mathrm{\mu m}$. Exposure time: 20 ms. Video recorded with a high-speed camera at 24 fps and played at recording speed. Multimedia view.}
\end{figure*}

\begin{figure*}[h!] 
\includegraphics[width=.48\textwidth]{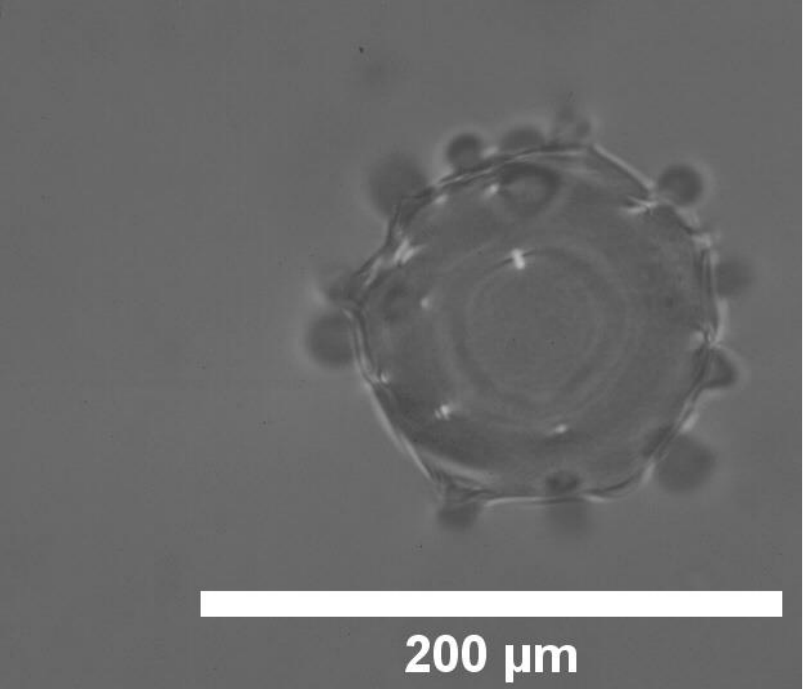}\hfill
\includegraphics[width=.48\textwidth]{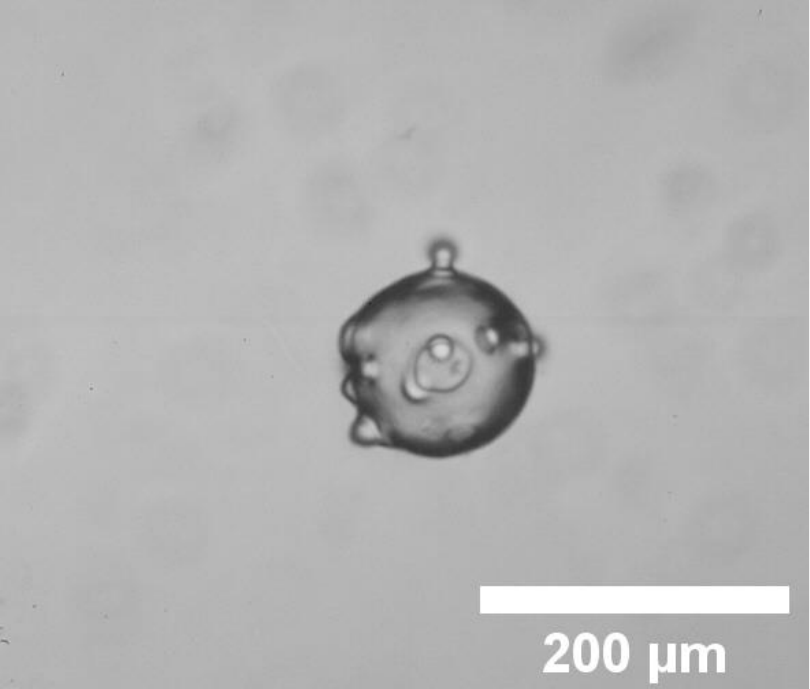}\hfill
\\[\smallskipamount]
\includegraphics[width=1.0\textwidth]{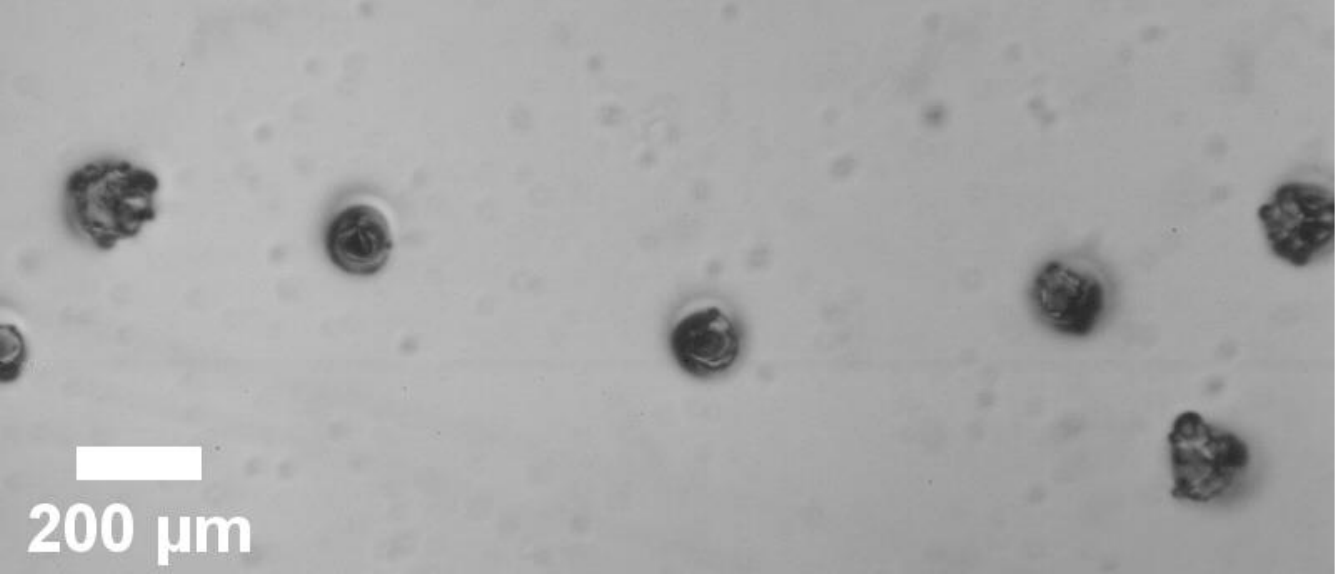}
\figuretag{S6}
\caption{\label{fig:Shrinking_Hair-dryer} Shrinking of PNiPAM mesogels suspended in a commercial Ludox suspension ($\varphi=0.350$) upon heating with a hair-dryer ($\dot{T}_{\mathrm{up}} \approx 35\mathrm{\degree C/min}$). Top left: magnification $\times 20$, field of view $277~\mathrm{\mu m} \times 238~\mathrm{\mu m}$, exposure time: 2 ms; video `FigS6left\_Mesogel\_shrinking\_Hair\_dryer\_x20.mp4' recorded at 10 fps and played at recording speed. Top right: magnification $\times 10$, field of view $711~\mathrm{\mu m} \times 533~\mathrm{\mu m}$, exposure time: 2 ms, video `FigS6right\_Mesogel\_shrinking\_Hair\_dryer\_x10.mp4' recorded at 10 fps and played at recording speed. Bottom: magnification $\times 3.5$, field of view $2133~\mathrm{\mu m} \times 917~\mathrm{\mu m}$, exposure time: 1 ms, video `FigS6bottom\_Mesogel\_shrinking\_Hair\_dryer\_x3.5.mp4' recorded at 10 fps and played at recording speed. Videos are available as independent supplementary material files. Multimedia views.}
\end{figure*}

When the variation in $T$ is quick, the shrinking process takes place in 3 steps.~\cite{mou_monodisperse_2014, sato_matsuo_kinetics_1988} (i) The quick initial shrinking leads to the formation of a dense skin layer at the surface of the mesogels. (ii) The skin layer is so dense that water can temporarily not diffuse out of the mesogels, causing their size to plateau. Meanwhile, the pressure inside the mesogels increases. (iii) When the inner pressure becomes high enough to overcome the strength of the skin layer, some areas of the skin layer are blown up like the surface of a balloon, forming bubble-like structures, and allowing water to be expelled from the mesogels as the expanded skin layer is no longer impermeable. Mesogel size further decreases. Note that, although commonly used in the literature by convenience,~\cite{mou_monodisperse_2014, sato_matsuo_kinetics_1988, okajima_kinetics_2002, kaneko_temperature-responsive_1995} the word `bubbles' is a somehow abusive as those actually correspond to pockets of water.

The 3-stage shrinking process described above can be seen in our videos of PNiPAM mesogels shrinking in a Ludox commercial suspension. Fig.~\ref{fig:Shrinking_3cpmin_x20} (Multimedia view) shows a frame taken from movie `FigS5\_Mesogel\_shrinking\_3Cpmin\_x20.mp4', available as an independent supplementary material file, where the mesogel is heated up at $\dot{T}_{\mathrm{up}} \approx 3 \mathrm{\degree C/min}$, \textit{i.e.} the same rate as that used in Figs 3, 4 (top part) and 5 (rightmost data points) of the main text. At stage (iii), the mesogel outer layer is subject to small local deformations and a few water pockets appear, grow and disappear over time. 
When the $T$ ramp is an order of magnitude faster ($\dot{T}_{\mathrm{up}} \approx 35 \mathrm{\degree C/min}$), the local deformations are significantly more pronounced, both in amplitude and in number, as seen in Fig.~\ref{fig:Shrinking_Hair-dryer} (Multimedia views), which shows mesogels shrinking in a commercial Ludox suspension recorded at various magnifications. In the associated movies (\textit{i.e.} `FigS6left\_Mesogel\_shrinking\_Hair\_dryer\_x20.mp4', `FigS6right\_Mesogel\_shrinking\_Hair\_dryer\_x10.mp4' and `FigS6bottom\_Mesogel\_shrinking\_Hair\_dryer\_x3.5.mp4', all available as independent supplementary material files), water escaping the mesogels can be visualized due to both the high speed of the flow and the difference between the water refractive index and that of the Ludox suspension the mesogels are suspended in. In the movie corresponding to the top-right panel of Fig.~\ref{fig:Shrinking_Hair-dryer} (`FigS6left\_Mesogel\_shrinking\_Hair\_dryer\_x20.mp4'), water release during stage (i) can also be observed. It appears as a thin light, even layer around the mesogel $\sim 6~\mathrm{s}$ after the beginning of the movie and is obvious after $\sim 9~\mathrm{s}$.

\section{\label{sec:DLS}Dynamic light scattering (DLS) data analysis}
\subsection{\label{sec:DLS_data_ROI}Selection of Regions of Interest (ROIs) for data processing}

\begin{figure*}[h!] 
\figuretag{S7}
\includegraphics{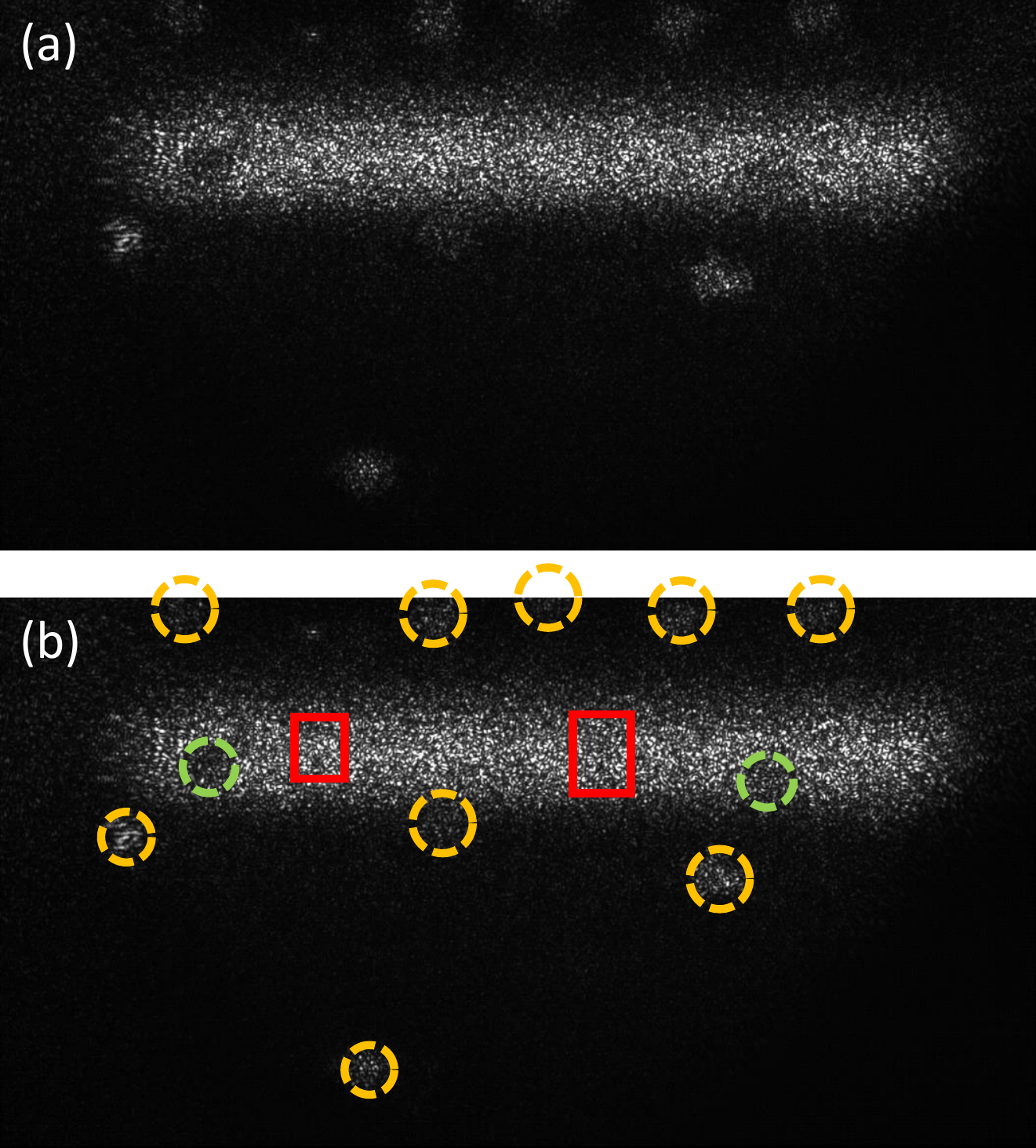}
\caption{\label{fig:ROI_selection} Images of the light scattered at 90 degrees by a Ludox-mesogels mixture at $39.5\mathrm{\degree C}$. The bright rod-shaped region near the top of the images corresponds to the scattering volume, \textit{i.e.} the sample region directly illuminated by the laser beam, whose thickness is $\sim 245~\mathrm{\mu m}$. (a) Raw image after the contrast has been enhanced. (b) Same image with annotations. The dashed circles are around PNiPAM mesogels. The mesogels appear darker than the background when they are either in the scattering volume or between the scattering volume and the camera (green circles), while they appear bright when located elsewhere (orange circles), see discussion in the text. Examples of two Regions of Interests (ROIs) selected to calculate the intensity autocorrelation function are shown as red rectangles.}
\end{figure*}

Fig.~\ref{fig:ROI_selection} shows a representative DLS image collected by the CMOS camera of our home-built set-up performing space-resolved DLS measurements. The rod-shaped region at the top corresponds to the scattering volume: it is bright because the laser beam illuminates the Ludox suspension, which scatters light collected by the camera objective lens. Some mesogels out of the scattering volume are also visible (highlighted by the orange circles in Fig.~\ref{fig:ROI_selection}(b)). Although they are not directly illuminated by the laser beam, these mesogels are visible because of scattering from the Ludox suspension. Indeed, the Ludox particles scatter light in all directions. Part of this light illuminates the mesogels located out of the scattering volume, which in turn scatter light that is collected by the set-up. Mesogels superposed to the scattering volume region appear as darker zones, highlighted by the green circles in Fig.~\ref{fig:ROI_selection}(b). They could be located either in between the scattering volume and the collection optics (partially blocking the light scattered by the Ludox suspension), or in the scattering volume (if the mesogels scatter less efficiently than Ludox at the detected scattering angle of $90$\degree).

To ensure we only probe light scattered by the Ludox suspension, we carefully checked all the images and processed selected areas of the scattering volume that were free of mesogels at all times during the measurements. Two such ROIs are shown as red rectangles in Fig.~\ref{fig:ROI_selection}(b).

\subsection{\label{sec:DLS_data_cream_conv}Creaming and convection at $T=35.9\mathrm{\degree C}$}
As mentioned when discussing Fig.~6 in Section~III of the main text, at $T = 35.9\mathrm{\degree C}$ mesogels moved in the sample. They initially underwent creaming, followed by convection after $\sim 4~\mathrm{h}$ as shown in the movies `NoFig\_creaming.avi' and `NoFig\_convection.avi' (available as independent supplementary material files). Videos were recorded at 1 fps and are played at 10 fps. The field of view is the same as that in Fig.~\ref{fig:ROI_selection}. Although we took care to process ROIs that were mesogels-free at all times, one may wonder if the motion of nearby mesogels due to creaming or convection may accelerate the Ludox dynamics, by providing some sort of `stirring' mechanism.

\begin{figure*}[h!] 
\includegraphics{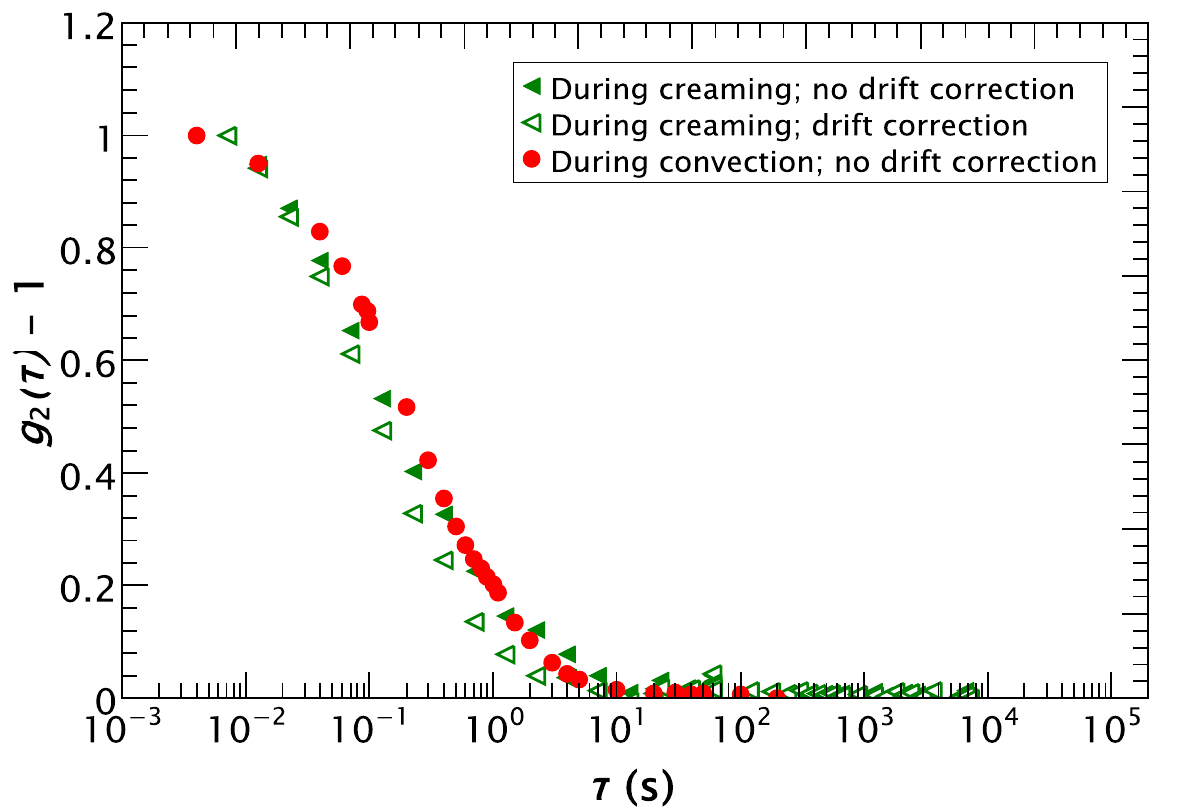}
\figuretag{S8}
\caption{\label{fig:IAC_cream_conv_drift} Intensity autocorrelation functions collected at $35.9\mathrm{\degree C}$, where mesogels exhibited creaming or convection. Data collected during creaming -- without and with drift correction, open and solid green triangles -- and convection, red circles, exhibit essentially the same behavior in spite of the difference of mesogel velocity for the two kinds of motion, supporting the argument that the Ludox dynamics are not affected by mesogel motion.}
\end{figure*}

To address this question, intensity autocorrelation functions (IACs) were calculated during both creaming and convection. Since convection involved mesogel motion at speeds much higher than for creaming (typically, $1.0~\mathrm{\um /s}$ \textit{vs} $0.18~\mathrm{\um /s}$), one would expect faster relaxation times during convection, had the mesogel motion had an impact on the Ludox dynamics. Fig.~\ref{fig:IAC_cream_conv_drift} shows that the Ludox dynamics is essentially the same, independently of the kind of mesogel motion. Furthermore, for the creaming phase, we also processed the data using the `drift correction' algorithm described in Ref.~\cite{cipelletti_simultaneous_2013}, which corrects the IACs for any spurious contribution due to a drift of the speckle pattern. We find very little difference between data with and without correction (compare the open and solid green triangles in Fig.~\ref{fig:IAC_cream_conv_drift}), ruling out any significant contribution of the mesogel motion to the measured Ludox dynamics. Finally, we recall that no mesogel motion was observed in all the other measurements shown in Fig. 6 of the main text.

\section{\label{sec:deltaphi}Calculation of $\varphi_L$, the effective volume fraction of the Ludox particles in the presence of mesogels}

In this section we shall derive expressions for the $T$-dependent volume fraction of mesogels and Ludox particles, as a function of the composition of samples prepared by mixing Ludox-alone and mesogel-alone stock suspensions. We start by calculating the effective volume fraction of the Ludox particles at a reference temperature, $T_{ref}$, defined as the temperature at which the two stock suspensions are mixed. To avoid any confusion, we use the subscripts `L' and `mgel' to designate the Ludox particles and the mesogels, respectively. Thus, $\varphi_L$ in this section correspond to $\varphi$ in the main text; it is defined as the volume occupied by the particles divided by the sample volume effectively available to them, \textit{i.e.} the total volume excepted that occupied by the mesogels:
\begin{equation}
\label{eq:phi_effective1}
\varphi_L(T_{ref}) = \frac{V_L \varphi_L^{(0)}}{V_L + V_{mgel}
\left[1-\varphi_{mgel}^{(0)}(T_{ref}) \left( \frac{d_{}(T_{ref})}{d^{(0)}(T_{ref})} \right)^3 \right]} \,,
\tag{S2}
\end{equation}
where $V_L$ and $V_{mgel}$ are the volumes of, respectively, the Ludox-alone and mesogel-alone stock suspensions mixed to obtain the final sample, and $\varphi_L^{(0)}$ and $\varphi_{mgel}^{(0)}$ are the corresponding volume fractions of Ludox and microgels in the two suspensions before mixing. The numerator of Eq.~\ref{eq:phi_effective1} represents the volume of the Ludox particles in the final sample and the denominator the sample volume accessible to them. $d$ and $d^{(0)}$ are the diameters of the mesogels when suspended in the final sample and in the medium of the mesogels-alone suspension (usually, water), respectively. Their ratio appears in the denominator of Eq.~\ref{eq:phi_effective1} as a correcting factor for $\varphi_{mgel}^{(0)}$, because the size of the mesogels is somehow smaller in a water-Ludox background as compared to pure water, as discussed in the main text. The ratio is raised to the third power because the volume occupied the mesogels scales as $d^3$.


By introducing the ratio $\chi = V_{mgel}/V_L$ of the volumes of the two initial suspensions and by taking into account the variation of $d$ (in the final suspension) with $T$, one derives the following expression for the Ludox effective volume fraction at any temperature:
\begin{equation}
\label{eq:phi_effective2}
\varphi_L(T) = \frac{\varphi_L^{(0)}}
{1 + \chi \left[1-\varphi_{mgel}^{(0)}(T_{ref})\left(\frac{d(T)}{d^{(0)}(T_{ref})}\right)^3\right]} \,.
\tag{S3}
\end{equation}

For completeness, we also report the volume fraction $\varphi_{mgel}$ of the mesogels in the final sample, calculated with respect to the total volume sample (\textit{i.e} including the volume of the mesogels themselves):
\begin{equation}
\label{eq:phi_mgel}
\varphi_{mgel}(T) = \frac{\varphi_{mgel}^{(0)}(T_{ref})\left[\frac{d(T)}{d^{(0)}(T_{ref})}\right]^3}
{1 + 1/\chi} \,.
\tag{S4}
\end{equation}
\begin{figure*}[h] 
\includegraphics{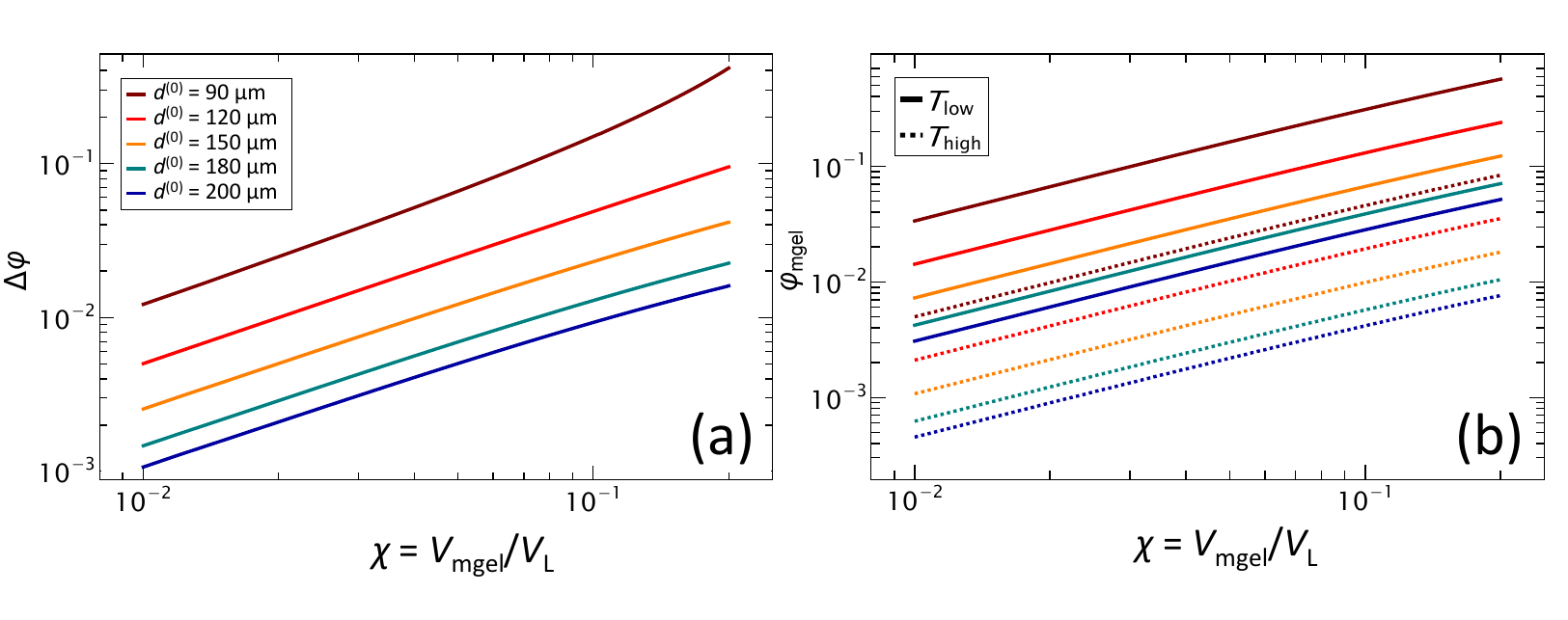}
\figuretag{S9}
\caption{\label{fig:phi_vs_chi} (a): Maximum achievable variation of the Ludox volume fraction, $\Delta \varphi$, and (b): the fraction of the total sample volume occupied by the mesogels, $\varphi_{\mathrm{mgel}}$, as a function of the ratio $\chi$ between the volumes of the two stock suspensions used to prepare Ludox-mesogel mixtures. Curves are obtained from Eqs.~\ref{eq:phi_effective2},~\ref{eq:phi_mgel} for (a) and (b), respectively, using the parameters for the mesogels fully characterized by optical microscopy: $\varphi_{\mathrm{mgel}}^{(0)} = 0.64$ and $\varphi_{\mathrm{L}}^{(0)} = 0.412$. The color code is identical for both (a) and (b) and refers to the diameter $d^{(0)}$ of the mesogels at the temperature at which the two stock suspensions are mixed together, as shown by the labels in (a).}
\end{figure*}

In practice, it is desirable to prepare samples that cover a broad range of dynamical behaviors upon changing $T$, while keeping the mesogel content as low as possible. To strike the right balance between these two conflicting requirements, we use Eq.~\ref{eq:phi_effective2} to determine the maximum achievable variation of Ludox volume fraction, defined as $\Delta \varphi = \varphi_{\mathrm{L}}(T_{\mathrm{low}}) - \varphi_{\mathrm{L}}(T_{\mathrm{high}})$, where $T_{\mathrm{low}}$ and $T_{\mathrm{high}}$ are the temperatures where the mesogel reach their largest and smallest size, respectively. We calculate $\Delta \varphi$ for a wide range of $\chi$ values, and for several values of $d^{(0)}(T_{\mathrm{ref}})$, to study the influence of the temperature at which the supernatant is removed from the mesogel suspension after it has been centrifuged and prior to adding the Ludox suspension.

Assuming that the mesogels in that suspension are randomly packed spheres, we use $\varphi_{mgel}^{(0)}=0.64$ in Eq.~\ref{eq:phi_effective2}. The results of these calculations are shown in Fig.~\ref{fig:phi_vs_chi}(a) for the mesogels synthesized according to the protocol of Section~\ref{sec:ball_prod_microscopy} and fully characterized with optical microscopy (see main text), using $\varphi_{\mathrm{L}}^{(0)} = 0.412$. As one may expect, at a given preparation temperature (\textit{i.e.} $d^{(0)}$ fixed), $\Delta \varphi$ increases when $\chi$ increases, which corresponds to the case where the number of added mesogels is increased while the volume of the Ludox suspension is kept constant or increased in lower proportion. At a given $\chi$ value, the smaller the mesogels in water (\textit{i.e.} the higher the preparation temperature $T_{\mathrm{ref}}$), the higher $\Delta \varphi$, because more mesogels are contained in a set volume of the mesogel stock suspension if the mesogels have a smaller size (recall that we assume a constant $\varphi_{mgel}^{(0)}=0.64$). For the same set of $d^{(0)}$ values, we plot in Fig.~\ref{fig:phi_vs_chi}(b) the fraction $\varphi_{\mathrm{mgel}}$ of the total sample volume occupied by the mesogels, calculated with Eq.~\ref{eq:phi_mgel} at both $T_{\mathrm{low}}$ (solid lines) and $T_{\mathrm{high}}$ (dotted lines). The curves shown in both panels of Fig.~\ref{fig:phi_vs_chi} provide guidelines for sample preparation. Importantly, they demonstrate that changes of a few \% of $\varphi_L$ (sufficient to vary the microscopic dynamics by several orders of magnitude for samples in the supercooled regime) are achievable even with less than 10\% of mesogels by volume.

\resub{As a final remark, we note that in experiments $T$ may slightly fluctuate. We use Eq.~\ref{eq:phi_effective2} to estimate the change $\delta \varphi_L$ of Ludox volume fraction due to a small temperature fluctuation $\delta T$, finding}
\begin{multline}
\label{eq:delta_phi_L}
\delta \varphi_L(T)\approx \frac{\partial  \varphi_L}{\partial T} \delta T =\\ 
= \frac{ \varphi_L^2 \delta T}{ \varphi_L^{(0)}} \left[ \chi  \varphi_{mgel}^{(0)}(T_{ref})
\frac{3 d^2(T)}{{d^{(0)}}^3(T_{ref})} \frac{\partial d}{\partial T} \right] \,.
\tag{S5}
\end{multline}
\resub{We evaluate $\delta \varphi_L$ for a temperature fluctuation $\delta T = 0.1~^{\circ}\mathrm{C}$ by inserting in Eq.~\ref{eq:delta_phi_L} the same set of parameters as for Fig.~\ref{fig:phi_vs_chi}, together with typical values $\chi = 0.05$, $\varphi_L = 0.38$ and $d^{(0)} = 90\um$. The largest fluctuation is found at intermediate temperatures, where the microgel size changes steeply with $T$: for $T = 31.3~^{\circ}\mathrm{C}$, $\delta \varphi_L = 1.4 \times 10^{-3}$, a rather small fluctuation. Far from the Volume Phase Transition Temperature, the impact of $T$ fluctuations is even lower: for $T = 24.5~^{\circ}\mathrm{C}$, $\delta \varphi_L = 4.5 \times 10^{-4}$, while at high temperature, $T= 35.3~^{\circ}\mathrm{C}$, $\delta \varphi_L$ is as low as $4.8 \times 10^{-7}$.}

\section*{Data Availability Statement}
The data that support the findings of this study are openly available in Zenodo at http://doi.org/10.5281/zenodo.5891491.

\nocite{*}
\bibliography{pnipam_balls}

\end{document}